\documentclass{emulateapj} 

\usepackage{natbib}
\usepackage{multirow}

\DeclareMathAlphabet{\mathsc}{OT1}{cmr}{m}{sc}
\def\testbx{bx}%
\DeclareRobustCommand{\ion}[2]{%
\relax\ifmmode
\ifx\testbx\f@series
                   {\mathbf{#1\,\mathsc{#2}}}\else
{\mathrm{#1\,\mathsc{#2}}}\fi
\else\textup{#1\,{\mdseries\textsc{#2}}}%
\fi}

\def\lya{\mbox{Ly$\alpha$}}

\shorttitle{A High S/N Composite Spectrum of GRB Afterglows}
\shortauthors{L. Christensen et al.}

\begin{document}
\title{A High Signal-to-Noise Ratio Composite Spectrum of Gamma-ray
  Burst Afterglows}


\author{L. Christensen\altaffilmark{1},
  J.~P.~U. Fynbo\altaffilmark{2}, 
  J.~X. Prochaska\altaffilmark{3}, 
  C.~C. Th{\"o}ne\altaffilmark{4}, 
  A. de Ugarte Postigo\altaffilmark{4},
  P. Jakobsson\altaffilmark{5}
}

\altaffiltext{1}{Excellence Cluster Universe, Technische Universit\"{a}t
        M\"{u}nchen, Bolzmanstrasse 2, 85748 Garching, Germany; lise.christensen@ph.tum.de}
\altaffiltext{2}{Dark Cosmology Centre, Niels Bohr Institute, University of 
     Copenhagen, Juliane Maries Vej 30, 2100 Copenhagen {\O}, Denmark}
\altaffiltext{3}{Department of Astronomy and Astrophysics, UCO/Lick
  Observatory, University of California, 1156 High Street, Santa Cruz,
  CA 95064, USA}
\altaffiltext{4}{INAF - Osservatorio Astronomico di Brera, via E. Bianchi 46, 23807, Merate, Lc, Italy}
\altaffiltext{5}{Centre for Astrophysics and Cosmology, Science Institute, University of Iceland, Dunhagi 5, IS-107 Reykjavik, Iceland}

\begin{abstract}
We present a composite spectrum of 60 long duration gamma-ray burst
(GRB) afterglows with redshifts in the range $0.35<z<6.7$ observed
with low resolution optical spectra. The composite spectrum covers the
wavelength range 700--6600 {\AA} in the rest frame and has a mean
signal-to-noise ratio of 150 per 1~{\AA} pixel and reaches a maximum
of $\sim$300 in the range 2500--3500~{\AA}. Equivalent widths are
measured from metal absorption lines from the \lya\ line to $\sim$5200
{\AA}, and associated metal and hydrogen lines are identified between
the Lyman break and \lya\ line.  The average transmission within the
Lyman forest is consistent with that found along quasar lines of
sight. We find a temporal variation in fine-structure lines when dividing the
sample into bursts observed within 2 hours from their trigger and
those observed later. Other lines in the predominantly neutral gas
show variations too, but this is most likely a random effect caused by
weighting of individual strong absorption lines and which mimics a
temporal variation.  Bursts characterized with high or low prompt GRB
energy release produce afterglows with similar absorption line
strengths, and likewise for bursts with bright or faint optical
afterglows. Bursts defined as dark from their optical to X-ray
spectral index have stronger absorption lines relative to the
optically bright bursts. The composite spectrum has strong
\ion{Ca}{ii} and \ion{Mg}{ii} absorption lines as commonly found in
dusty galaxies, however, we find no evidence for dust or a significant
molecular content based on the non-detection of diffuse interstellar
bands.  Compared to starburst galaxy spectra, the GRB composite has
much stronger fine-structure lines, while metal absorption lines are
weaker.

\end{abstract}

\section{Introduction}
Optical spectra of long duration gamma-ray burst (GRB) afterglows
provide a means to study the physical condition of the interstellar
medium in their host galaxies through the associated metal absorption
lines. While the host galaxies themselves are generally too faint to
allow for detailed spectroscopic studies, spectra of the afterglows
obtained shortly after the burst can achieve much higher signals. High
resolution spectra of individual bursts have shown that GRBs reside in
galaxies with large gas column densities and sub-solar metallicities
\citep{vreeswijk04,chen05}. Host galaxy abundance ratios and
metallicities have been measured \citep{savaglio03,vreeswijk06}, and
evidence for galaxy outflows has been established from the presence of
multiple absorption components separated in velocity space
\citep{schaefer03} and from correlation of absorption line species
with host galaxy luminosity \citep{chen09}. The distances from the
burst site to the absorbing material has been determined to be larger
than a few tens of pc \citep{dessauges06,prochaska06b,chen07,delia10},
while in other cases the time variations of fine-structure lines
\citep{vreeswijk07,delia09,ledoux09}, or the presence of vibrationally
exited H$_2$ molecules \citep{sheffer09} indicated a distance of
0.4--2 kpc. While this is in itself interesting for the study of the
nature and physical conditions present in high redshift star forming
galaxies, one of the goals of afterglow absorption line studies is to
constrain the nature of the progenitor of the GRB, but the diverse
properties of GRB afterglows have made this difficult.

At the average burst redshift of $\langle z \rangle \sim 2.2$
\citep[][see also Web site\footnote{\tt
    http://www.raunvis.hi.is/$\sim$pja/GRBsample.html} with an updated
  sample]{jakobsson06a} observed with the \textit{Swift} satellite
\citep{gehrels04}, the rest-frame UV spectra reveal that GRBs occur in
galaxies which contain large amounts of neutral gas. Generally, the
hydrogen column density determined from the GRB spectra lie in the
damped Lyman-$\alpha$ (DLA) regime
log\,($N$(\ion{H}{i})/cm$^{-2}$)~$>$~20.3, and both hydrogen and metal
column densities are typically larger than observed in intervening
DLAs towards quasars \citep{savaglio03,jakobsson06b}. Quasar sight
lines with DLAs intersect and probe the interstellar environment of a
galaxy at an impact parameter which is typically less than 10~kpc
\citep[e.g.,][]{moeller02,moeller04}, although some candidates DLA
galaxies are found at even larger impact parameters
\citep{christensen07} in the outskirts of the galaxies. In contrast,
long duration GRBs explode in young star forming galaxies
\citep[e.g.,][]{christensen04b}, and are located close to the most
luminous region in the host \citep{bloom02a,fruchter06,svensson10}.
The difference between the two samples can be described by the
diversity in impact parameters \citep{prochaska07,fynbo08,pontzen10}.

While high spectral resolution studies can reveal the impact on the
environment near the GRB event, low spectral resolution studies
obtained for fainter afterglows also provide valuable information.
However, significant discrepancies have been found between GRB
afterglows where high resolution spectra seem to indicate lower
metallicities than the afterglows observed with lower resolution
\citep{ledoux09}, although the effects or biases which cause this
discrepancy are not understood.

The current knowledge of the interstellar medium (ISM) in GRB host
galaxies relies on a few well studied afterglows at a range of
redshifts. Additional knowledge can be gained by taking advantage of a
large set of GRB afterglows at different redshifts which sample
different rest-frame wavelengths. By creating a composite spectrum, we
can study the typical properties in a high quality spectrum over a
large rest-frame wavelength range. Here we present a composite
spectrum of 60 GRB afterglows from the largest sample which has been
analyzed in a consistent manner \citep{fynbo09,fynbo10}. The composite
spectrum allows us to measure which lines are typically present in a
GRB afterglow and their average equivalent widths and metallicities.

This paper is organized as follows.  In Sect.~\ref{sect:relevance}
  we explain the benefits and limitations for the physical
  interpretation of the composite afterglow spectrum.  We describe
the full sample and how the composite spectrum is created in
Sect.~\ref{sect:sample}. In Sect.~\ref{sect:results} the global
properties are investigated and the list of absorption lines is
presented, while in Sect.~\ref{sect:absorp} we investigate specific
absorption lines in detail. In Sect.~\ref{sect:lbg} the absorption
line strengths are compared with other high redshift
galaxies. Metallicities are derived in Sect.~\ref{sect:coldens}. We
discuss the results and potential bias effects in the interpretation
of the composite spectrum as well as future prospects in
Sect.~\ref{sect:disc}.


\section{Motivation for constructing a composite spectrum}
\label{sect:relevance}

The goal of this study is to use a unique database of low-resolution
spectra to explore whether fainter lines are detectable when combining
all spectra compared to detections in individual, lower
signal-to-noise (S/N) spectra. Since all afterglow spectra seem to be
unique, we also want to explore specific absorption line features and
their possible dependence on global properties for the GRB
explosion. A detailed investigation of the variations of each detected
absorption line in the set of spectra will be presented in a
forthcoming paper (A. de Ugarte Postigo et al., 2011, in preparation).

A potential benefit of having a high signal-to-noise ratio afterglow
composite spectrum which covers a large rest-frame wavelength range is
to compare existing or future individual low-resolution afterglow
spectra to answer a set of questions: (i) Does the (new) optical
afterglow spectrum have unusual absorption lines or ratios thereof?
(ii) What absorption lines are detectable given the GRB redshift and
the S/N of the data?  (iii) Is it possible to derive redshift of a
very noisy afterglow spectrum by convolving it with the composite
spectrum?

Certain limitations exist for the analysis of a composite spectrum
too, where some quantities are better explored with high resolution
optical spectra. First, the velocity spreads of individual
absorption lines are not possible to analyze at low spectral
resolution.  Second, some lines may be particularly blended with
other lines, for example molecular H$_2$ lines which fall in the
$\lya$ forest. In addition, iron fine-structure lines are affected by
blending in the low resolution data, so a detailed analysis of their
temporal variations has to use higher spectral resolution data.

Individual afterglows have shown a wide range of metallicities in GRB
hosts, and the composite is biased towards the brightest afterglows in
the full sample as explained in Sect.~\ref{sect:sample}. However, with
this in mind we can use the wide rest-frame wavelength coverage of the
composite spectrum combined with the detection of a large number of
weak absorption lines, to derive the average column density of a large
number of metal species in Sect.~\ref{sect:coldens}. This presents a
unique possibility for the study of the interstellar medium (ISM) in
faint, high redshift galaxies, although the diversity of metallicities
probed by GRBs have to be investigated using individual higher
resolution spectra.

\section{Afterglow sample}
\label{sect:sample}
The largest systematic analysis of low resolution optical spectra of
long duration GRB afterglows performed to date is presented in
\citet[][hereafter F09]{fynbo09}. While the majority of spectra in F09
involves low-resolution spectra, they also include five high resolution
spectra observed either with UVES on the VLT or HIRES at the Keck
Telescope. To date, optical spectra of GRB afterglows are mostly
obtained at low resolution because afterglows are typically faint
($R>20$ mag) at the time of the start of the observations.  The low
spectral resolution data from the sample in F09 provide the basis for
this present follow-up study.

\subsection{Sample selection}

The aim of the sample selection is to construct a sample of
long GRBs that is selected independent of the optical properties
of the afterglows and at the same time has as high completeness
in high quality optical follow-up as possible.

To get a well selected sample which is independent of the optical
properties of the afterglow, and to achieve a high degree of
completeness and good quality optical spectra, a set of selection
criteria was used according to the methods described in
\citet{jakobsson06a}: (i) the burst duration is longer than 2 sec.,
(ii) accurate localization obtained with X-ray observations, (iii)
small foreground Galactic extinction ($A_V\le0.5$~mag) in the
direction of the burst, (iv) declination useful ($-70^{\circ}<\delta <
+70^{\circ}$) for optical follow up observations, and finally (v) a
solar distance greater than 55$^{\circ}$ at the time of the GRB. A
total number of 146 \textit{Swift} bursts observed between March 2005
and September 2008 fulfill these criteria, and optical spectra have
been obtained for 69 of these (see F09 for details on each individual
object).

From the full sample of spectra, we here focus on the 66 afterglows
which were observed with low-resolution optical spectroscopy using the
following instruments: FORS1 and FORS2 on the VLT, AlFOSC on the
Nordic Optical Telescope, GMOS-N and GMOS-S on the Gemini North and
South, respectively, and finally LRIS on the Keck Telescope. Depending
on the grating or grism chosen for the observations, these instruments
provided spectra with resolutions between 315 and 2140. We exclude
afterglow spectra with uncertain redshift determination due to the
lack of absorption lines in their spectra, and where the redshifts
were determined from photometric redshift methods from their host
galaxies. We also exclude afterglow spectra where only upper limits on
the redshifts have been established.  In total 60 afterglow spectra
pass these criteria and were used to create the composite
spectrum. Their redshifts lie in the range $0.35<z<6.7$, with a mean
of $z=2.22$.

Given the variety of instruments used, time after the trigger for the
spectroscopic observations following the burst, and intrinsic
magnitude of the afterglow, the S/N of the individual spectra vary
between a few and $\sim$200 for the burst with the highest signal
(GRB\,060729 at $z=0.5428$).

For the $z\gtrsim2$ GRBs, F09 determine their neutral hydrogen column
densities to lie in the range
$17.0<$~log\,($N$(\ion{H}{i})/cm$^{-2}$)~$<22.7$, and most of them are
in the DLA regime. Metallicities have been determined for a subsample
of afterglows which give typical values between 1\% and 30\% solar
\citep{watson06,fynbo06,prochaska07,fox08,eliasdottir09,prochaska09}. The
associated metal absorption lines are typically very strong and are
certain to be associated with the host galaxy itself.

Although the observational biases for the sample as a whole are well
known (F09), the effects on the composite spectrum can be severe. For
example, dust obscuration will render the afterglow fainter, and the
S/N will typically be lower.  Bursts which are observed early, have
low extinction, lie at lower redshifts, occur in a certain
environment, or also possibly in an environment with a specific
chemical composition at a fortuitous distance from the burst, are more
likely to have a bright afterglow.  Because of the weighting scheme
for creating the composite spectrum these bursts will dominate over
the remaining ones. Alternative combination schemes are found to
produce a spectrum with a worse S/N and one with a smaller useful
wavelength range, but general properties are found to be consistent,
so this paper focuses mainly on the weighted combination scheme
described below.

\subsection{Composite spectrum}

Each individual spectrum was normalized with a cubic spline
function. The continuum level was estimated by fixing spline points in
regions that were not contaminated with any absorption lines, either
intrinsic to the GRB or from intervening systems as listed in the
tables in F09, or from the atmosphere. The regions for the continuum
was estimated by eye, and special care was taken in the wavelength
regions around and blueward of the \lya\ line.  This manual procedure
was estimated to be most appropriate since the individual spectra have
widely different quality and spectral range, so that an automatic
procedure would be very complicated. Since the normalization was done
by eye, the errors introduced on a pixel-by-pixel scale are not known,
so we cannot propagate the error from the normalization procedure
throughout. With iterations of the normalization, we estimate from the
spectra with the best signal that we can achieve an accuracy better
than 1\% for the normalization. Whenever possible, we used the flux
calibrated afterglow spectra to derive the normalized spectra, because
these should follow a pure power law. For 15 of the 60 spectra, we did
not have a flux calibrated spectrum, while other flux calibrated
spectra had artificial features introduced by errors in the flux
calibration from the spectrophotometric star as mentioned in
F09. These features were removed too by the normalization.

The associated error spectra were normalized as well using the same
spline function.  GRB afterglows can have different spectral slopes,
so the normalization was a necessary step before the composite
spectrum is computed. Any specific signature of the GRB environment
such as dust extinction, and the 2175 {\AA} bump that appeared for
example in the \object{GRB\,070802} afterglow
\citep{kruehler08,eliasdottir09}, is eliminated during this
process. Here we do not aim to analyze the composite, line of sight
averaged, extinction curve.

Before combining the spectra, we masked absorption features that were
unrelated to the GRB environments. To avoid contamination by
absorption lines from intervening absorption systems, we masked each
of the detected lines arising in intervening systems which were not
associated with the GRB (see Tables 5--72 in F09). We masked fringes
present at red wavelengths in AlFOSC spectra, and fringes present at
wavelengths larger than 7000~{\AA} in spectra obtained with the blue
sensitive FORS1 detector installed in April 2007. We also masked
atmospheric features such as the strong telluric bands around 6900,
and 7600 ~{\AA}, but also weaker telluric bands around 6300, 7200,
8200, and 9000 {\AA} were masked; if not, they would introduce
artifacts and spurious absorption lines redward of 3000~{\AA} in the
rest frame. The masking was done by assigning a large value ($10^6$)
to the error spectrum at the corresponding wavelengths.

The typical dispersion obtained with the FORS V300 grating is
3.25~{\AA} pixel$^{-1}$ and the average redshift of all the bursts
which contribute to the composite is $z=2.22$. The average redshift
therefore matches the dispersion of the low resolution perfectly
\(\lambda_{\rm rest}=\lambda_{\rm obs}/ (1+z\)), so the optimal choice
  is to shift each spectrum to $z=0$ with a dispersion of
  1\,{\AA}~pixel$^{-1}$.  We investigated the outcome of the composite
  spectrum by choosing other resampling sizes for the dispersion,
  e.g., 0.5 and 0.8 {\AA} pixel$^{-1}$, however, the absorption lines
  were not better resolved and the S/N decreased correspondingly. We
  adopted the redshift for each GRB from neutral species associated
  with the GRB environment as listed in the tables in F09.

The spectra were combined using the weighted average
\begin{equation}
f_{\lambda,\mathrm{comb}}= \sum_i (f_{i,\lambda}/\sigma^2_{i,\lambda})
/  \sum_i 1/\sigma^2_{i,\lambda}
\end{equation}
where the flux $f$ at each wavelength $\lambda$ was weighted by its error
$\sigma$, and summed over all spectra, $i$. The associated
error spectrum was also calculated:
\begin{equation}
\sigma^2_{f,\lambda,\mathrm{comb}}= (\sum_i 1/\sigma^2_{i,\lambda})^{-1}.
\end{equation}

The resulting composite spectrum and its error spectrum is shown in
Fig.~\ref{fig:comp}, and Table~\ref{tab:composite_spec} lists the
relative fluxes and uncertainties. The median value of the error
spectrum is 0.0065 from 1000--2000~{\AA}, corresponding to a median
S/N~=~150. At wavelengths 2000--3500{\AA} the error is smaller, and
the S/N~=~250--300 per {\AA}, as illustrated in the middle sub-panels
in Fig.~\ref{fig:comp}. At wavelengths larger than 3500~{\AA} the S/N
depends on the exact region, where only few, low signal data
contribute, and where some regions have lower signal because the input
spectra receive a low weight in regions of strong telluric absorption
lines.  Only one low redshift afterglow (GRB\,061021 at $z=0.3463$)
contributes to the composite beyond 5800~{\AA}, and here the noise is
significantly larger than in other spectral bins.

Apart from the regular error spectrum there are other effects that
add to the noise. The composite spectrum is affected by uncertainties
from the normalization of the individual spectra. If this is not done
correctly, we can artificially introduce absorption lines in the
composite. At the very high S/N level, it is necessary that the
normalization is done to a level better than 0.5--1\%, otherwise
features which resemble absorption lines will be introduced to the
composite spectrum artificially and the uncertainty will introduce
systematic errors for the equivalent widths.  The normalization is
particularly important for the individual spectra with good S/N which
dominate the composites, however, when the S/N is good, the continuum
flux level is also easier to identify visually.

As afterglow spectra show a range of abundances and equivalent widths
for individual absorption lines, it is interesting to determine the
spreads of values with wavelength. We examined the variance of values
from the sub-set of the 60 spectra that went into each wavelength
bin. This variance spectrum essentially shows a flat behavior with a
value around 0.05--0.1 from 1000-4000{\AA}, while redward of 4000
{\AA} it has a larger scatter of values between 0 and 0.5. The reason
why the variance spectrum does not have a larger value at the location
of the metal absorption lines is likely because many of the individual
spectra have low S/N, so they will dominate the spreads of values. One
exception exists at the \lya\ line around 1216--1230{\AA}, where the
variance is larger (0.3--0.4) because of the large range of hydrogen
column densities probed by individual GRBs.

\begin{figure*} 
\figurenum{1}
\plotone{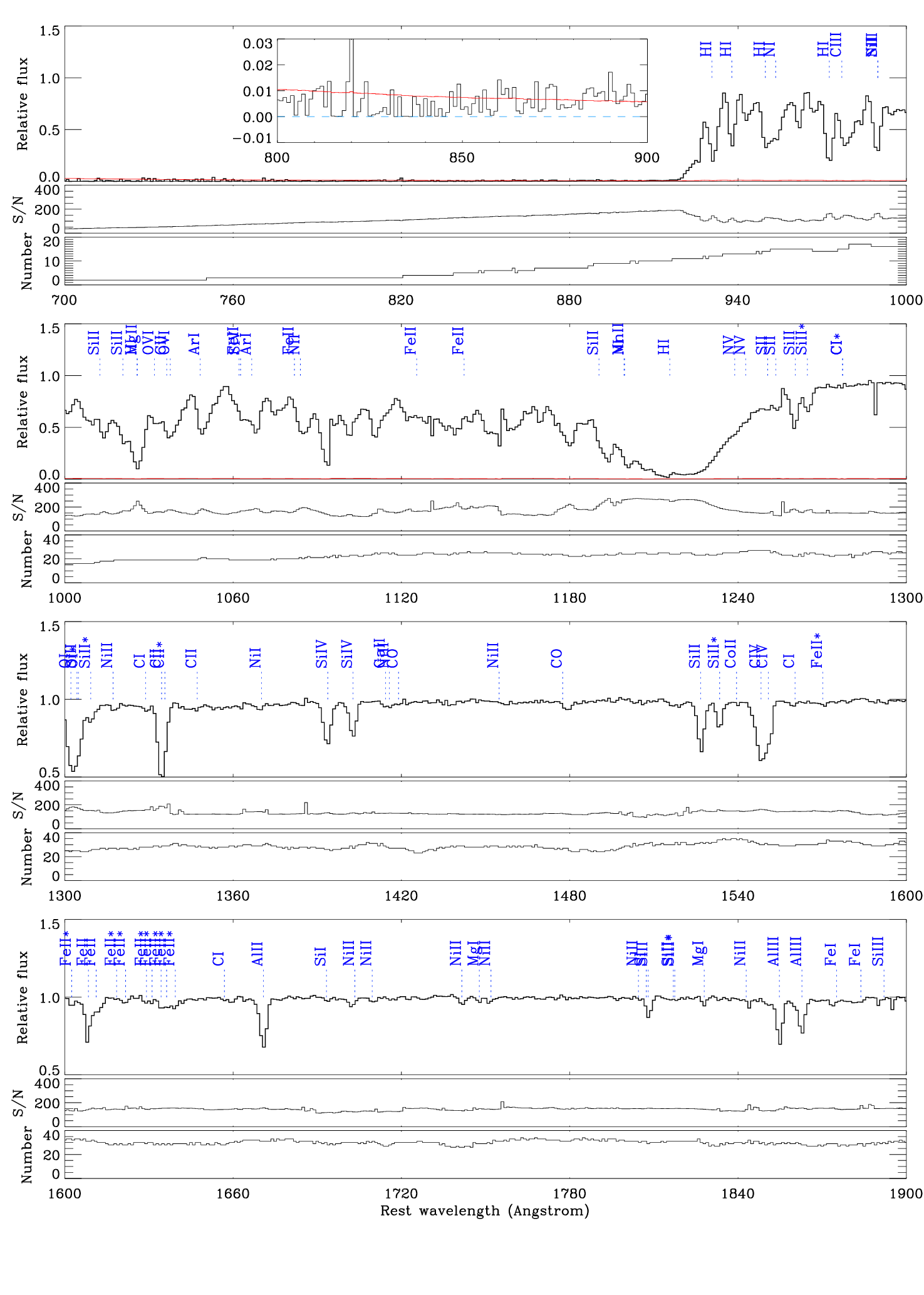} 
\caption{Each sub-spectrum consists of three parts.  \textit{Top:}
  The composite spectrum and its associated error spectrum (red line)
  which typically has a value around 0.005. \textit{Middle:} The
  corresponding S/N per pixel, and the \textit{Bottom:} shows the
  number of afterglow spectra that are used in the composite per
  spectral pixel. All detected absorption lines from
  Tables~\ref{tab:abslist} and \ref{tab:bluelist} are indicated. Note
  the change of scale on the y-axis, which implies that the error
  spectrum is not always visible.}
\label{fig:comp}
\end{figure*}
\begin{figure*} 
\figurenum{1} 
\plotone{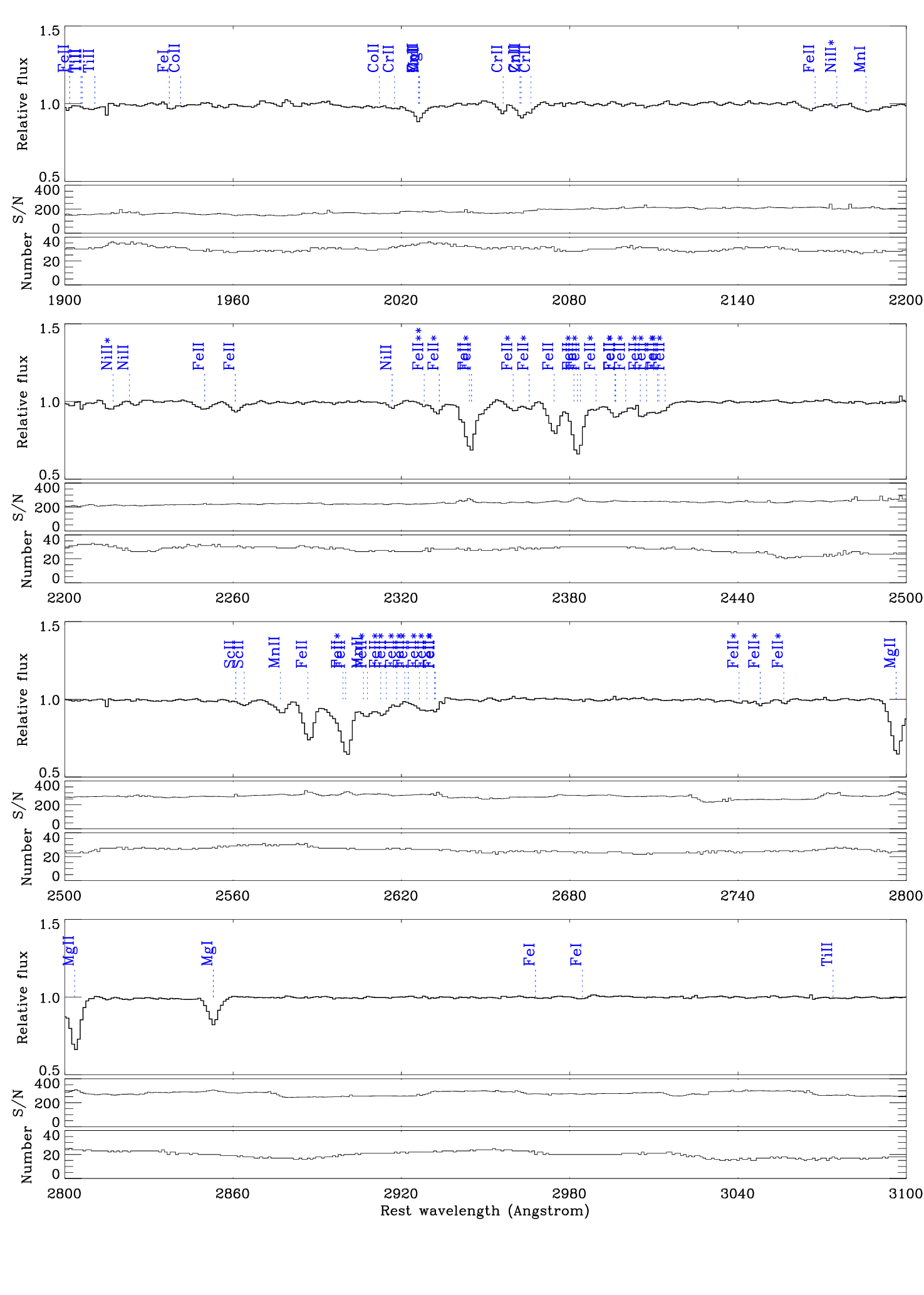} 
\caption{\textit{Continued.}}
\end{figure*}
\begin{figure*} 
\figurenum{1} 
\plotone{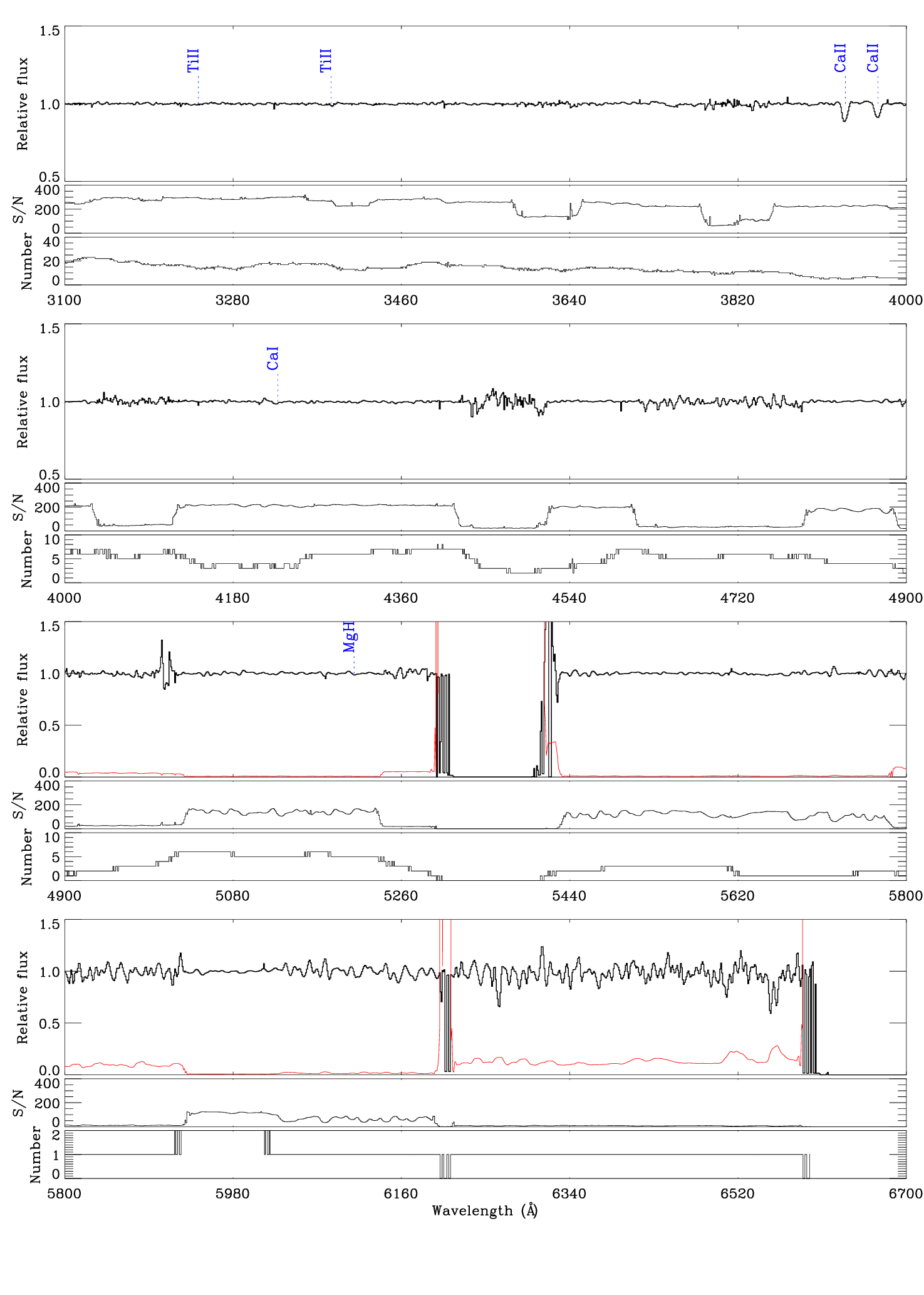} 
\caption{\textit{Continued.}}
\label{fig:comp}
\end{figure*}

A few individual high S/N spectra will necessarily dominate the
composite spectrum. In order to check for systematic effects we also
created a simple median spectrum, again masking out intervening
absorption systems as well as atmospheric absorption lines.  The
median spectrum has a lower S/N~$\sim$~50, estimated by the standard
deviation in regions free of absorption lines, and a useful spectral
range between 1000--2700~{\AA}.

In addition to this, we also investigated the output where the spectra
were multiplied with each other and divided by the number of systems.
In practical terms, the spectra were added in logarithmic space, while
omitting the masked pixels.

\begin{equation}
\log(f_{\lambda,\mathrm{comb}})= \Sigma_i \log (f_{i,\lambda})/N_{\mathrm{\lambda}},
\end{equation}
where $N_{\mathrm{\lambda}}$ is the number of spectra which contributes
to each wavelength bin. Without the normalization, the composite
spectrum will be completely dominated by absorption lines, and have a
very low output S/N. 

Physically, this method corresponds to placing the absorbing systems
one after the other along the line of sight, and then taking the
average.  In this case we avoid the systematic effects from weighting
the individual spectra, however, we are again limited by the resulting
S/N which is also $\lesssim$50. This implies that many of the faintest
absorption lines are lost in the noise.

\section{Results}
\label{sect:results}
Unless specifically mentioned, the following sections involves the
weighted combination of the 60 afterglows, because we found that the
general results did not depend on which of the three combination
methods we used. The largest amount of information was therefore
obtained with the highest S/N spectrum with the largest rest-frame
wavelength coverage. Some results were also derived by creating sub
samples from the full sample of bursts.

\subsection{Absorption line list}
Compared to individual spectra, we see significantly more absorption
lines in the composite. Even the high resolution afterglow spectra
obtained, e.g., with UVES or HIRES, are not able to detect some of the
fainter lines at a high significance; although high resolution spectra
are better suited to investigate lines \citep[e.g.,][]{ledoux09} which
are blended at the lower resolution.

Table~\ref{tab:abslist} lists all the lines and rest-frame equivalent
widths ($W_r$) detected at above the 3$\sigma$ uncertainty level.
Among all the lines identified, we checked if there are any missing
lines that should be detectable. From the singly ionized species which
are the dominant ionization state, we only find three non-detections,
all of which are expected to be very weak. The non-detection of
\ion{Ni}{ii} $\lambda$1502, \ion{Co}{ii} $\lambda$1466, and
\ion{Cr}{ii} $\lambda$2040 are therefore not surprising.  Some
afterglow spectra have unidentified absorption lines, as several ones
in F09 do. After a careful masking of telluric features in the
individual spectra, the composite spectrum does not appear to have any
unidentified lines. Other very weak absorption lines which were
uniquely detected in the GRB\,080607 spectrum, such as \ion{O}{i}
$\lambda$1355, \ion{Mg}{i} $\lambda$1683, \ion{Co}{ii} $\lambda$1574
and \ion{Ge}{ii} $\lambda$1602 \citep{prochaska09}, are not recovered
in the composite spectrum.

Identified lines which have rest-frame wavelengths separated by
$2-5$~{\AA} were deblended by fitting Gaussian functions to the blends
of lines using the {\tt splot} routine in {\tt IRAF}, keeping the
wavelengths of the lines fixed, and fitting a single absorption line
width. This routine also estimated the uncertainties using a
representative value from the composite uncertainty spectrum around
the line blend to represent the pixel statistics. Lines closer
together than $\sim$2~{\AA} were not deblended and here the total
$W_r$ is listed. Reference rest-frame wavelengths for the transitions
were found in the atomic table in \citet{morton03}. The $W_r$
uncertainties do not include the uncertainty for the placement of the
continuum.

Since some spectra with high S/N dominate the composite, we compared
with the median composite and the multiplied spectra. Although their
overall S/N are lower, we still find the same absorption lines in the
three composites and the $W_r$ of the strong absorption lines are
consistent to within the errors. 

To check further whether some of the weaker lines are randomly caused
by errors in the normalization of the individual spectra, we created
two independent composites. The spectra were sorted according to their
S/N, and two subsets of composite spectra with the same S/N were
created.  Since the GRBs have different redshifts, the best method
would be to make two composites with a similar S/N ratio throughout
the spectral range, and with a similar redshift distribution so that
there are an equal number of objects contributing to each spectral
pixel. This is not feasible in practice, so the spectra were simply
combined according to their S/N ratio, and the two spectra have almost
the same average redshift $z=2.4$ and $z=2.0$, respectively.  We
checked whether the absorption lines found in the full composite
spectrum are present in both of the sub composites. This is not always
the case, as noted in the comments to Table~\ref{tab:abslist}. We
cannot rule out that the lines are not real based on this test; the
absorption lines may potentially be present in very few lines of
sight, but randomly one that was observed at a high S/N. Additionally,
the presence or absence of the absorption lines can be due to temporal
variations as discussed in Sect.~\ref{sect:temporal}. In particular,
some lines appear quite strong in one composite, but are absent in the
other composite spectrum.

Due to the wide range of $N$(\ion{H}{i}) column densities, the
\lya\ absorption line does not have a purely damped line profile, and
residual emission can be seen in the \lya\ trough. Nevertheless, the
average value of log\,$N$(\ion{H}{i}) = 22.0 provides a good
Voigt profile fit to the red wing of the line in the composite
spectrum. At this high column density, the red wing of the absorption
line extends to 1400 {\AA} which is consistent with the theoretical
line profile. It also affects the continuum estimate of the metal
absorption lines in this wavelength range.

Blueward of the DLA, a mixture of metal absorption lines and
intervening hydrogen absorption lines are present. Hence, measurements
of the equivalent widths of lines in this region are likely to be
contaminated by intervening \ion{H}{i} clouds. Identified lines with
wavelengths $<$~1215~{\AA} are listed in Table~\ref{tab:bluelist}.

As a consequence of the low resolution of the spectra, we see no
evidence for any velocity shift of the absorption lines relative to
each other or multiple components of individual lines as reported in
studies of individual afterglows observed with higher resolution
spectroscopy \citep[e.g.,][]{moeller02,fiore05,chen05,prochaska08,fox08}.

\subsection{\lya\ forest transmission}
At $z\sim2$, the intergalactic medium (IGM) expresses itself in quasar
spectra as a series of \ion{H}{i} Ly$\alpha$ transitions commonly
termed the Ly$\alpha$ forest \citep{rauch98}.  In a single spectrum, the
Ly$\alpha$ forest appears as a series of discrete absorption lines.
If one considers an ensemble of sightlines, however, the stochastic nature
of absorption produces an average, effective opacity $\tau_{\rm eff}$.
Using high S/N spectra of bright quasars
\citet{dallaglio08} fits the transmission in the \lya\ forest as
a function of redshift by the equation \(T= e^{-\tau_{\mathrm{eff}}}\) where \(
 \tau_{\mathrm{eff}}= 10^{-2.21\pm0.09}(1+z)^{3.04\pm0.17}\). In
contrast to quasars which ionize their surroundings and create a proximity
region with higher transmission nearby, GRBs and their
hosts are expected to have relatively less impact on their
surroundings, and the proximity effect will not be as
pronounced. Therefore, GRBs are excellent probes of the transmission in
the \lya\ forest, in particular, in the view that they can be observed
at extremely high redshifts where the current record holder is
GRB\,090423 at $z=8.2$ \citep{tanvir09,salvaterra09}.

At the average redshift of $z=2.22$ for the composite spectrum, the
expected transmission in the \lya\ forest is $T=0.8\pm0.1$.  In the
composite spectrum, many of the absorption lines in the \lya\ forest
are associated with the ISM in the GRB host galaxy (see
Table~\ref{tab:bluelist}). Without correcting for these lines, we find
that the transmission is 0.6, but if we avoid the associated lines,
the regions of the largest transmission where no metal absorption
lines are expected suggests $T\sim0.85$, as illustrated in the first
two panels in Fig.~\ref{fig:comp}.  This is in excellent agreement
with the prediction from the quasar transmission.  Thus, we find no
overall difference in the \lya\ forest transmission towards GRBs and
quasars.  To analyze the forest and its evolution in further detail,
one needs to look at higher spectral resolution data.

\subsection{Lyman limit emission}

The existence of the Ly$\alpha$ forest demonstrates that the IGM is
predominantly ionized at $z<6$.  While quasars are known sources of
ionizing radiation, they may not dominate the extragalactic UV
background, especially at $z>3$ \citep[e.g.,][]{faucher-giguere09}.
Instead, star-forming galaxies may contribute the majority of ionizing
photons, but this requires that they have a non-negligible escape
fraction $f_{\rm esc}$. Since GRB hosts are potentially unbiased
tracers of star formation they have been used as probes of the
ionizing radiation escaping blueward of the Lyman limit into the
IGM.  The estimated escape fraction from the
\ion{H}{i} column density distribution function for GRB afterglows at
$z>2$ is $\langle f_{\mathrm{esc}}\rangle$ = 0.02$\pm$0.02 \citep[][;
  F09]{chen07b}.  This value is consistent within the 1$\sigma$ error
with numerical simulations, which predict an escape fraction of 1\%
\citep{pontzen10}.

Blueward of the Lyman limit in the composite spectrum, we measure a
relative flux level of 0.0060$\pm$0.0050 in the wavelength range from
800 to 900 {\AA} as illustrated in the small inserted panel in
Fig.~\ref{fig:comp}. At these rest-frame wavelengths, only 11
afterglow spectra contribute to the composite, and some only cover
redward of 880 {\AA} and have different S/N ratios in this wavelength
range as listed in Table~\ref{tab:lylimit}. Most afterglows at $z>2$
have \ion{H}{i} column densities in the DLA regime, where the escape of
ionizing radiation is effectively blocked. Indeed, the average
\ion{H}{i} column density from the bursts in Table~\ref{tab:lylimit}
is $10^{22.1\pm0.1}$~cm$^{-2}$, and at this high value the radiation
blueward of the Lyman limit is completely absorbed.

The average redshift for those afterglow which contribute blueward of
the Lyman limit is $z=3.66$. At this redshift the average transmission
in the \lya\ forest is $T=0.52$ \citep{dallaglio08}, and in the
Ly$\beta$ forest it is $T=0.57$ \citep{madau95}. Including these two
transmission corrections, the measured relative flux blueward of the
Lyman limit is $f=0.020\pm0.017$, which is consistent with the values
derived from the \ion{H}{i} column density distribution.

Since clouds with column densities in the DLA regime effectively
absorb all the Lyman limit photons, a significant escape fraction is
only expected to be present for low column density systems.  A few
afterglows have revealed column densities well below the DLA limit,
and GRB\,050908 with log\,($N$(\ion{H}{i})/cm$^{-2}$)=17.6$\pm$0.1 has a
significant level of ionizing radiation detected beyond the Lyman
limit ($f_{\mathrm{esc}}=13\pm4$\%), and appears to be a unique object
among the whole sample. Because of the relatively low S/N for the
GRB\,050908 spectrum, eliminating it from the total composite changes
flux blueward of the Lyman limit very little: \(f =
  0.018\pm0.017\). A couple of other systems observed with low
  resolution spectroscopy (GRB\,060124, GRB\,070411, and GRB\,071020)
  have relatively low \ion{H}{i} column densities, but their spectral
  ranges do not reach the Lyman limit. More studies which cover the
  very blue end of the spectral range are needed to determine the
  effects of the ionizing radiation from GRB host galaxies, and their
  impact on the extragalactic UV background radiation field.

\subsection{Temporal variations}
\label{sect:temporal}
Our composite spectrum ignores the case of time variations in the fine
structure levels. Several spectra show that the absorption lines
strengths can change with time due to the temporal variation of the
ionizing radiation from the central explosion
\citep{dessauges06,vreeswijk07,delia09,ledoux09}. 

The spectra included in this investigation are obtained at one
particular time (Table 2 in F09) at an average of 3 hours after the
burst in the rest frame for the composite spectrum. Since the spectra
contribute to the composite weighted by their signal, the effective
rest-frame observing time can be calculated by weighting according to
a representative value for the S/N of each spectrum. This gives an
effective observing time of $t_{\rm rest}=2.78$ hours after the burst
in the rest frame for the composite spectrum.  We investigate the
cumulative effect of the observing time on the spectra, and look for
temporal variations in absorption lines. We create two subsamples
separated according to the time of the start of the spectroscopic
observations after the burst trigger. The division takes into account
the redshift of the GRB by shifting the start of the observing time to
the rest frame. These sub-samples of composite spectra include 30
afterglow spectra obtained within 1.9 hours from the trigger (in the
rest frame of the burst) which has an average redshift of $z=2.4$, and
30 spectra obtained between 1.9 and 14 hours, which have an average
redshift $z=2.0$. Both composite spectra have a S/N between 100 and
200 in the wavelength range 1500--3500\,{\AA}. The magnitudes of the
afterglows at the time of the trigger vary by $\sim$~4 mag in both the
early and late-time samples, while the early sample is on average 1.4
mag brighter than the late sample (See Fig. 2 in F09).

\begin{figure*} 
\figurenum{2} 
\plotone{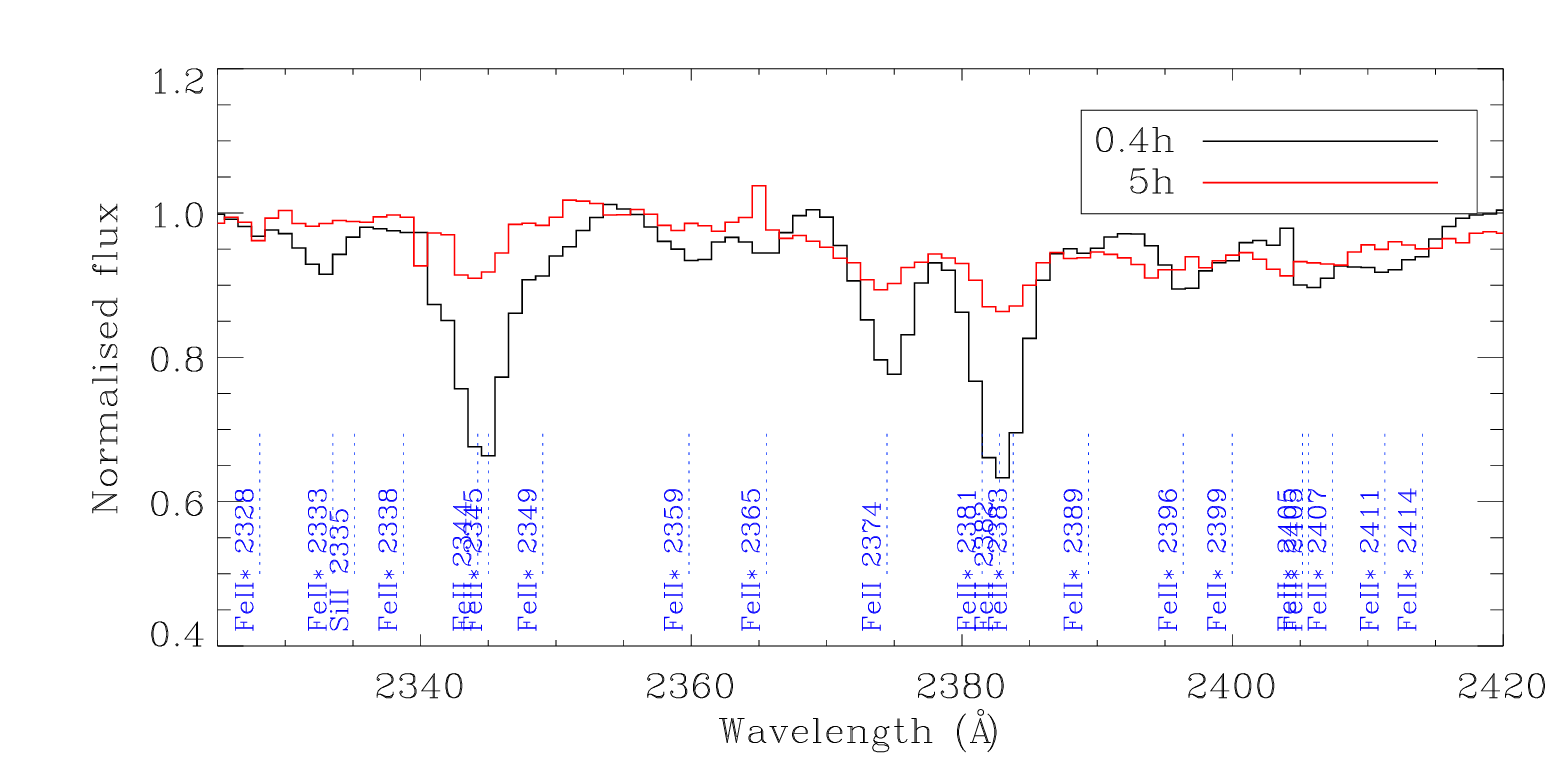} 
\caption{Apparent variation of the absorption line strength around the
  wavelength region where strong \ion{Fe}{ii} and \ion{Fe}{ii}$^*$
  lines dominate. The spectra are labeled according to the average
  time for starting the spectroscopic observation after the trigger
  (0.4 and 5 hours, respectively). Strong variations of the fine
  structure lines are seen, but variations of the ground states
  (\ion{Fe}{ii} $\lambda$2344, 2374, 2382) are most likely caused by
  the sampling of the spectra and is not physically real.}
\label{fig:temporal}
\end{figure*}

\begin{figure*} 
\figurenum{3} 
\plotone{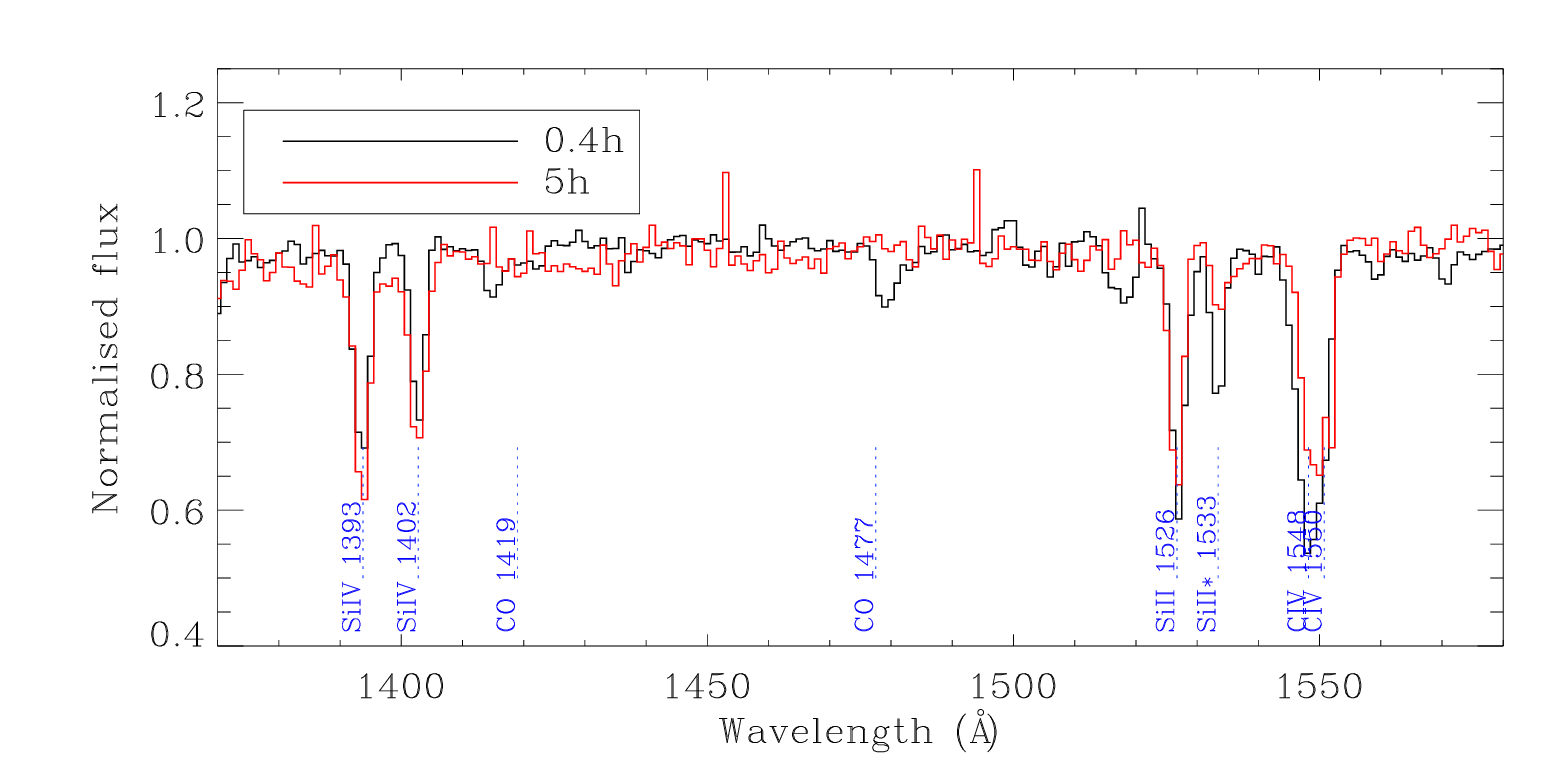} 
\caption{The composites show smaller variations within 20\% for the
  high ionization lines \ion{C}{iv}\,$\lambda\lambda$1548,1550 and
  \ion{S}{iv}\,$\lambda\lambda$1393,1402. Note the disappearing CO
  lines and weakening of the \ion{Si}{ii}* line. }
\label{fig:temporal2}
\end{figure*}

A clear difference in the strength of the metal absorption lines is
seen. Strong variations in the iron fine-structure lines are observed
from the spectrum obtained at an average of 0.4 hours after the burst
to that obtained 5 hours after the burst. However, also the species
\ion{Fe}{ii}, \ion{Al}{iii}, \ion{Mg}{ii}, \ion{Zn}{ii}, \ion{Cr}{ii},
\ion{Ni}{ii} which trace the neutral gas seem to decrease in
strength. In particular, the \ion{Fe}{ii} $W_r$ decrease by a factor
of 2--4 as illustrated in Fig.~\ref{fig:temporal}.  Between the two
composites, all the absorption lines appear to decrease in strength by
a factor of $\sim$2. Other lines like \ion{C}{ii}, and \ion{Si}{ii}
and the high ionization lines \ion{Si}{iv} and \ion{C}{iv} remain
constant within 10--20\% as illustrated in Fig.~\ref{fig:temporal2}.
The disappearing CO lines in Fig.~\ref{fig:temporal2} could also be
ascribed to a temporal variation, but as explained in
Sect.~\ref{sect:molecules} the CO lines are only detected in the
afterglows from GRB\,080607 and possibly also GRB\,060210. Hence, it
is not certain whether this is a real effect or just a coincidence
from the strong CO lines in one GRB sight line. It is curious that
GRB\,080607 also has the earliest rest-frame observation time in the
entire sample.

The simple median spectrum and the multiplied spectrum composites do
not suffer from the possible biasing effects of weighting with noise
level. Nevertheless, they also reveal changes in the absorption line
strengths (both ground and excited states) between the early and the
late time samples.

\begin{figure*} 
\figurenum{4} 
\plotone{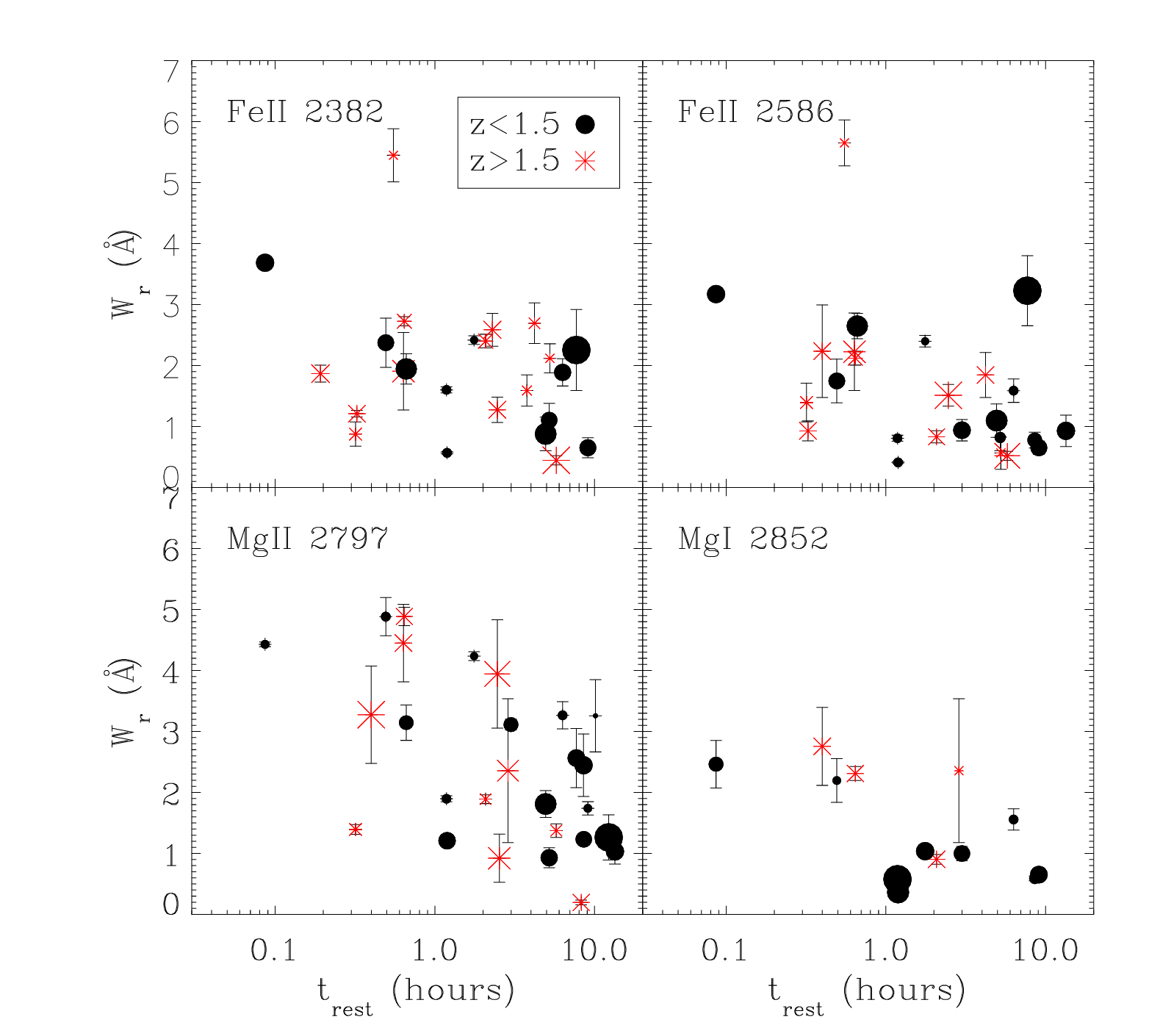} 
\caption{Rest-frame equivalent widths measured for individual bursts
  at $z<1.5$ (\textit{dots}) and $z>1.5$ (\textit{crosses}). The
  symbol sizes scale with the logarithm of the S/N of the individual
  spectra. Both early and late time GRBs have spectra with a good S/N.}
\label{fig:ewplot}
\end{figure*}

To date, no time series of individual afterglows have revealed any
significant variations in the non-exited states. In particular, a time
series of high resolution spectra of GRB\,071031 showed constant high
ionization lines \citep{fox08}. Therefore, we investigate whether
other effects are at play in the composite spectra, and find the
temporal changes of absorption line strengths to be a random effect
caused by some afterglows with very strong absorption lines. In order
to conclude this, we investigate $W_r^{\ion{Mg}{ii}}$ and
$W_r^{\ion{Fe}{ii}}$ from individual systems as a function of the
starting time for the spectroscopic observations as well as a function
of redshifts as shown in Fig.~\ref{fig:ewplot}. First, there are no
systematic changes of the $W_r$ with redshift. Second, there is a
tendency that earlier observations have larger $W_r$, but including
the measured $W_r$ uncertainties any statistically significant
correlation disappear. Both the early- and the late-time observations
have spectra with high S/N ratio.  Taking the $W_r$ at face value in
addition to the fact that the earlier observations have brighter
afterglows and therefore give more weight to the composite, this
mimics the behavior of a temporal variation of the composite
spectrum. We therefore suspect that this effect is not physically
real. A similar analysis of the \ion{C}{iv} and \ion{Si}{iv} $W_r$
measurements and their S/N show a large scatter of values at all
times, so the apparent constant value in the composites is also likely
to be a random effect.

Apart from the composition of the external medium, the fine-structure
line variations also depend on the input UV flux from the GRB. We
investigated the effect by analyzing subsets of composites by
separating according to the spectral slope $\beta$, derived from the
power-law dependence of the flux $f_{\nu}\propto\nu^{-\beta}$. Values
of $\beta$ are listed for half of the bursts in \citet{kann10}, while
the remaining ones are obtained by fitting a power law to the flux
calibrated spectra without including additional extinction. Including
a set of possible extinction laws allows the spectral index a large
range of values within the uncertainties. Here, we use the face values
without including the uncertainties and create two subsamples
separated at $\beta=1$. Comparing the composite spectrum with lower
$\beta$ values (brighter UV than optical flux in $f_{\lambda}$ units)
with that obtained from afterglows with $\beta>1$, we find that both
have strong fine-structure lines, with no clear trend that one has
systematically much stronger lines than the other. A more careful
treatment of the spectral slope dependence would need to include
systematic analyses including the uncertainties and the intrinsic
extinction to derive the intrinsic UV flux from the afterglows. Such
an analysis is beyond the scope of this paper.

\subsection{Effect of GRB energy release}

A trend between the GRB isotropic energy release and the afterglow
brightness has been found \citep{kann10}, albeit with a very large
scatter possibly due to variations in the circumburst density. Because
of this possible connection, we explore the effects for the composite
spectrum by the GRB energy release. GRBs and their afterglows are
thought to be collimated with a typical beaming angle between
1--25$^{\circ}$ \citep{frail01}.  Out of the 60 afterglows
investigated here, 42 bursts have a measured bolometric isotropic
energy \citep[][and references therein]{kann10}.  These 42 bursts have
energies between $51.38 < \log E_{\mathrm{iso,bol}} \mathrm{(erg)}<
54.49$.

We investigated the effects for the composite
spectrum by dividing the sample into two samples according to the
estimated bolometric, isotropic energy release from the prompt
emission.  Each subsample contains 21 GRBs. The energy at which
the sample is divided is determined by statistical sampling rather
than a physically real property. The low and high energy composites
have $51.38 < \log E_{\mathrm{iso,bol}} \mathrm{(erg)}< 52.88$ and
$52.88 < \log E_{\mathrm{iso,bol}} \mathrm{(erg) }< 54.49$,
respectively. Overall, there is no significant difference between the
absorption line widths for either the high or low-ionization lines,
and fine-structure lines are present in both the high-energy composite
and low-energy composite. However, redward of 2200~{\AA}, all lines
(Fe, Mn, Mg) in the low-energy composite are a factor of $\sim$4
weaker.  The reason for this is most likely due to the fact that a
couple of high-energy GRB afterglows at $z\sim1$ (GRB\,061121 and
GRB\,080413) have rather strong Fe absorption lines and dominate the
absorption lines at rest-frame wavelengths larger than
2200~{\AA}. Neglecting these two GRBs, the high- and low-energy
composite spectra are similar, and the variations between the lines
widths are 11\%$\pm$22\% with no systematic trends of which of the low-
or high-energy composites produce the strongest lines. We conclude
that the prompt GRB energy release itself has no effect on the
absorption lines.

\subsection{Dark bursts}
For some bursts, afterglows are not detected, but the success of
detecting the afterglows depend strongly on the afterglow search
strategies, i.e., rapid searches which are able to probe faint
magnitudes. Some bursts may be dark, because they lie at high
redshifts and are therefore not detected in optical follow-ups. Others
may be obscured by dust. A better method to define intrinsically dark
bursts relies on the X-ray flux versus the simultaneous optical
afterglow magnitude. From this optical to X-ray spectral index,
\citet{jakobsson04} define dark busts as those having
$\beta_{\mathrm{OX}}<0.5$.

\begin{figure*} 
\figurenum{5} 
\plotone{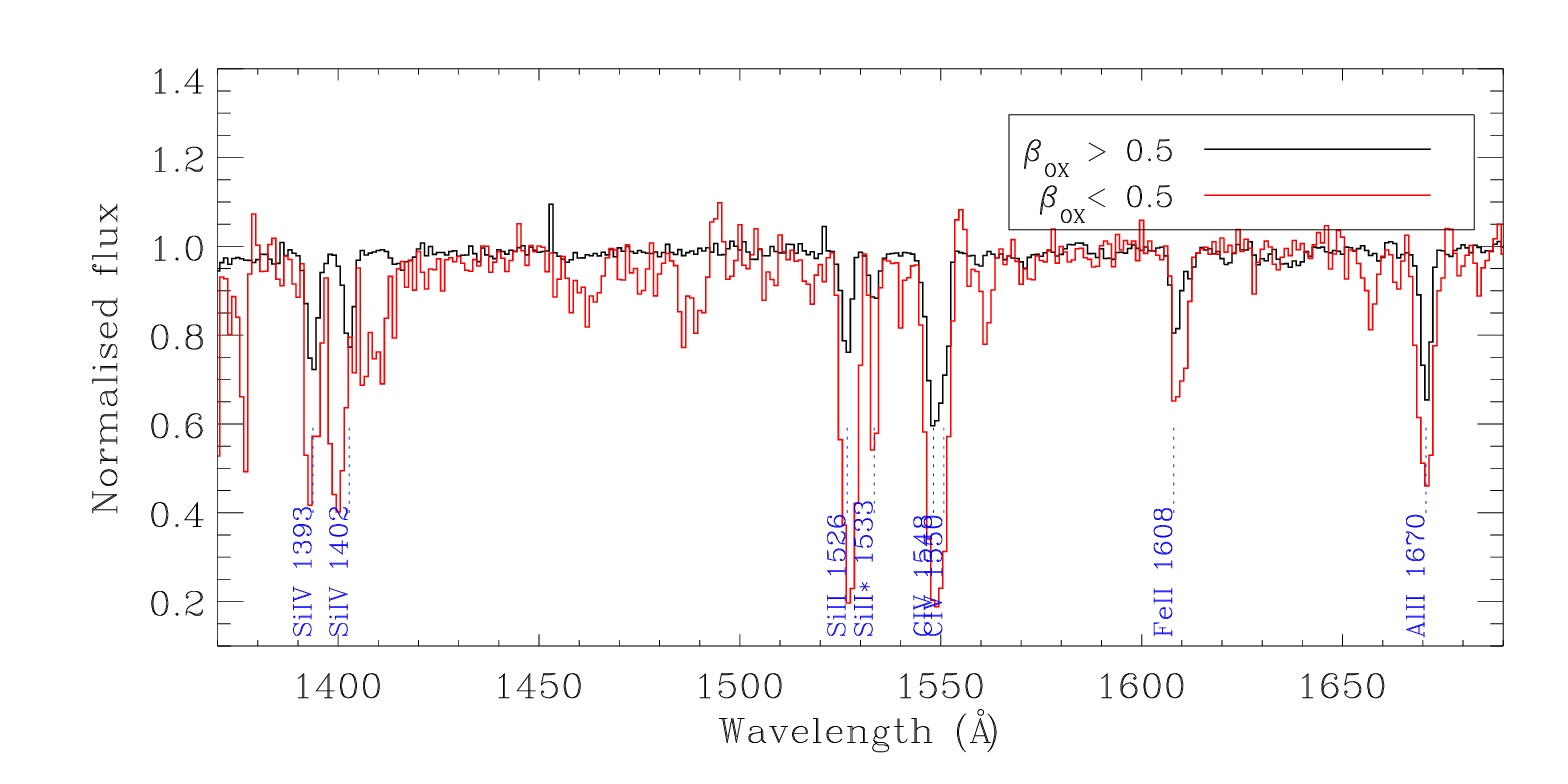} 
\caption{Dark bursts with optical to X-ray spectral
  slopes $\beta_{\mathrm{OX}} < 0.5$ have stronger low ionization,
  high ionization and fine-structure absorption lines compared to the
  optically brighter ones with slopes $\beta_{\mathrm{OX}} >0.5$. }
\label{fig:darkburst}
\end{figure*}

In the sample of 60 afterglows, 53 have measured $\beta_{\mathrm{OX}}$
(Table 73 in F09), and 10 of these satisfy the criterion for being a
dark burst. We created composite spectra for those with
$\beta_{\mathrm{OX}}<0.5$ and $\beta_{\mathrm{OX}}>0.5$, but because
of the large difference in the number of spectra that go into the
composite, the S/N is also unequal. The composite spectrum from the
bright burst sample with $\beta_{\mathrm{OX}}>0.5$ has an S/N between
100 and 200 from 1000 to 3000~{\AA}, while the composite from the dark
burst sample with $\beta_{\mathrm{OX}}<0.5$ has a S/N~=~25--40 in the
same wavelength region. Still, the dark burst composite has stronger
absorption lines by a factor of 2--3 over this spectral range, while
\ion{H}{i} column density decreases by 0.2 dex from the bright to the
dark composite. A representative wavelength region is shown in
Fig.~\ref{fig:darkburst}. 

While the optical to spectral index separation at $\beta_{\rm OX}=0.5$
is based on a physical reasoning, we may still wonder if artifacts are
included from comparing two spectra with very different S/N
ratios. Choosing instead a different separation of $\beta_{\rm OX}$
for the subsamples may involve bursts with slopes that are dependent
on the particular positions of the spectral breaks between the optical
and X-ray bands. Although the separation is not arising from a
physical explanation, the low-$\beta_{\rm OX}$ sample will still
include the 'darkest' bursts. Dividing the full sample into two equal
size parts instead with a separation in the spectral index of
$\beta_{\mathrm{OX}}=0.75$ such that the S/N is comparable,
consistently shows that the darker bursts have stronger absorption
lines. The detection of stronger metal absorption lines towards dark
bursts is fully consistent with the idea that these reside in more
metal-rich, dustier ISM relative to the bright counterparts. This is
already established from analysis of the properties of the host
galaxies from broad band photometry \citep{perley09} and from X-ray
absorption (F09).

\subsection{Afterglow optical luminosities}
The optical luminosities of the afterglows depend on a large number of
parameters, including the density of the external medium. Although the
medium which causes the absorption lines in afterglow spectra is
generally determined to lie at distances greater than 1 pc, we may
still investigate if the composite spectrum shows any changes related
to the optical luminosities. Potentially, some absorption lines may
arise closer to the GRB explosion than otherwise determined, or there
may be an anti-correlation between a more dusty, metal rich
environments and the optical luminosity of the afterglows.

Having investigated the effect of the spectral slopes, we include an
analysis of any systematic effects of the afterglow optical
brightness. Analyzing the light curves in a systematic way,
\citet{kann10} determine the observable, extinction corrected
magnitudes of a large number of afterglows at $t=1$ day from the
explosion in the rest frame of the bursts.  The $R_C$ magnitudes are
listed for 39 of the 60 bursts in the composite sample. We divide the
spectra into two almost equal sized subsamples separated at
$R_C=19.55$ mag. The two resulting composites have the same S/N ratio.

The optically faint subsample has stronger \ion{Fe}{ii} lines
redward of 2200~{\AA} compared to the bright subsample. While the
high ionization lines in the two samples are of the same strength, the
low-ionization ones are about 50\% stronger in the faint sample. As
for the isotropic energy release case, two bursts (GRB\,061121 and
GRB\,080413) are among the optically faint ones 1 day after the
explosion. At the time when the spectra were obtained they were among
the brighter ones. Eliminating these two GRB afterglows from the faint
sample, the composite spectra have similar absorption line strengths.

\section{Metal absorption lines}
\label{sect:absorp}
In this section, we compare the individual features in the composite
spectrum with other categories of absorption line systems with the
purpose of investigating whether GRB environments are special. This is
interesting in the context of constraining the nature of the GRB
progenitor through any peculiarities in the surrounding medium.

\subsection{Strong \ion{Ca}{ii} and \ion{Mg}{ii} absorbers}
The composite spectrum displays a large number of absorption lines. To
compare the composite with other high redshift galaxy spectra, we are
limited to the stronger transitions found in large samples in low
spectral resolution data, because these are the ones most easily
detected in faint galaxy spectra.

The \ion{Ca}{ii} absorption lines are very strong in the composite
with $W_r^{\ion{Ca}{ii}~\lambda{3934}}$~=~0.77 {\AA}, which places the
average GRB site in the category of the dustiest absorbers
\citep{zych09} compared to other intervening absorptions systems
selected by their neutral hydrogen column densities. In intervening
quasar absorbers at lower redshifts, strong \ion{Ca}{ii} lines belong to
DLAs with a color excess of $E_{B-V}>0.1$ mag \citep{wild06}. The
reverse is not true; only 10\% of DLAs have strong \ion{Ca}{ii}
lines \citep{nestor08}, and DLA selected absorbers have an order of
magnitude less reddening \citep[$E_{B-V}<0.02$~mag
  in][]{murphy04}. 

Also the \ion{Mg}{ii} absorption lines fall among strong, but not
extreme absorbers detected in quasar spectra \citep{nestor05}. Its
$W_r^{\lambda 2797}=1.7$~{\AA} is common for systems associated with
DLAs \citep{rao06}.  The \ion{Mg}{ii} $W_r$ is smaller than the
$W_r=3.8$~{\AA} found in a composite spectrum of local starburst
galaxies (C. Tremonti, private communication) as well as ultra strong
($W_r>3.0$ {\AA}) \ion{Mg}{ii} absorbers at $z=0.7$ \citep{nestor10},
which possibly trace starburst galaxies \citep{rubin10}. The widths of
\ion{Mg}{ii} lines in quasar spectra which arise in intervening
galaxies at $0.3<z<1$ also depend on the galaxy's rotation and
inclination \citep{kacprzak10}. Since GRB hosts have low stellar
masses \citep{savaglio09}, star burst driven galactic winds would
easily escape their potential well. Strong lines with large $W_r$ such
as \ion{Si}{ii} $\lambda$1526 have very large optical depths such that
the line widths measure the velocity spread of individual components
in DLA systems both towards quasars and GRBs
\citep{prochaska08b}. Large widths can therefore be seen as proxies of
galaxy wind strengths, which would indicate that GRB hosts have weaker
winds relative to galaxies with higher absolute star formation rates
(SFRs).  Higher resolution spectra are needed to explore the
  effects of galaxy winds in GRB hosts in order to quantify the effect
  the host mass and SFRs on the absorption line widths of the very
  strong lines.

\subsection{Dusty afterglow spectra}

There is currently little evidence either for a strong extinction from
the afterglows themselves with typical values of $E_{B-V}<0.3$
\citep{galama01,kann06}, while a few individual bursts suffer from more
extinction like GRB\,070306 which had $A_V=5.5\pm0.6$ mag
\citep{jaunsen08} and GRB\,070802 which had $A_V=0.8-1.5$ mag and
showed a pronounced 2175 {\AA} bump \citep{eliasdottir09}. These two
GRB spectra have relatively low S/N, so they do not affect the
composite significantly. Recently, a very high extinction of $A_V>12$
was suggested for the dark burst GRB\,090417B
\citep{holland10}. However, such dusty environments are rare in the
sample studied here, where the extinction measured from the afterglows
is generally low \citep{kann10}.

In the integrated spectral energy distribution (SED) of the host
galaxies there is evidence for relatively little reddening around
$E_{B-V}\sim0.1$ \citep{christensen04b,savaglio09,levesque10},
although fits to SEDs which include the far-IR emission find larger
amounts of extinction in GRB hosts \citep{michalowski08}. Analyses of
individual afterglow spectral slopes generally only allow for small
amounts of extinction in the burst region \citep {kann10}. If no dust
is present in the host galaxy, this would produce a brighter optical
afterglow and introduce an observational bias \citep{fynbo01a}. For
some GRBs no optical afterglow is detected \citep{jakobsson04}, which
is most likely due to dust obscuration in their host galaxies
\citep[][; F09]{perley09}.  However, for the composite spectrum of
bursts with detected optical afterglows, dust obscuration is not a
strong effect.

\subsection{Molecular lines and diffuse interstellar bands}
\label{sect:molecules}
We search the composite for absorption lines arising from molecular
lines, but only CO lines are found. These are very rare in GRB sight
lines, and the composite is dominated by the strong CO lines observed
in GRB\,080607 and which are caused by excitation from the GRB
\citep{prochaska09,sheffer09}. A possible, but weaker detection is
that towards GRB\,060210 (F09). Eliminating the GRB\,080607 afterglow
while creating the composite, the two detected CO lines in the
composite vanish (see Fig.~\ref{fig:temporal2}).  Regarding other
molecular lines, H$_2$ absorption lines are not commonly found in GRB
afterglows \citep{tumlinson07,ledoux09}. The strong H$_2$ lines
observed in the line of sight to GRB\,080607 are blended with too many
other lines in the composite \lya\ forest, so we cannot include a more
detailed study here.

We also search for diffuse interstellar bands (DIBs) redward of
4000~{\AA} using a list of DIBs found in the Galaxy
\citep{jenniskens94}. In particular, we look for the strongest bands
at $\lambda$4428, $\lambda$5705, $\lambda$5780, $\lambda$5797,
$\lambda$6010 $\lambda$6203, and $\lambda$6283, which apart from seen
in Galactic sight lines, are also found in nearby starburst galaxy
spectra. Some of these lines have recently been observed in a
high-resolution spectrum of the Type Ib SN\,2008D which exploded in
NGC 2770 \citep{thoene09}, showing that DIBs can be present in
galaxies which form massive stars.  These lines are found to have
typical strength of $>0.1$~{\AA}, and their equivalent widths
correlate with the color excess $E_{B-V}$ and the hydrogen column
density \citep{heckman00,welty06}. Even considering the very high
$N$(\ion{H}{i}) in the composite, we failed to get any significant
detection although the spectral range and S/N ratio for the composite
is sufficient to find some of the bands with
$W_r\approx0.1$~{\AA}. \citet{welty06} derive a best-fit relation
between the 5780 {\AA} DIB strength and the reddening in Galactic
sight lines. Using their fit \(\log E_{B-V} \approx -2.7 + 1.1 \log
  W_R(5780) \)/m{\AA}, the limit for $W_r$ corresponds to an upper
 limit of $E_{B-V}<0.3$ mag. A similar reddening limit would be inferred by a UV-shielded line of sight in the Small Magellanic Cloud \citep{cox07}.

Relative to stars in the Galaxy, weaker DIBs are found in the less
metal-rich medium of the Magellanic Clouds \citep{welty06}. A higher
ionizing radiation field than that present in the Galaxy may be
responsible for destroying the carriers of the molecules that give
rise to the DIBs \citep{welty06,cox07}. Apart from detections of DIBs
in the Galaxy, DIBs have been found in some gas-rich
galaxies. \citet{lawton08} find that strong DIBs are absent in six out
of seven DLAs at $z<0.5$, with the one detection belonging to the most
metal rich system with the highest reddening
\citep{junkkarinen04}. Like the DLA systems, the average GRB host ISM
is also metal-poor and contain relatively little dust.  Additionally,
GRB regions are known to have strong radiation fields
\citep{levesque10}, and therefore the DIB carriers are likely to be
destroyed in the region near the GRBs.  These arguments combined may
be the reason for the non-detection of DIBs in the composite
spectrum.

 The most commonly found DIBs fall in the rest frame visible region of
 the composite spectrum which has a significantly smaller S/N ratio,
 and where only few afterglows contribute. It would be more useful to
 look for DIBs in the UV part of the spectrum, which allows the
 detection of weaker lines. Some unidentified UV features with
 equivalent widths of a few to a few tens of m{\AA} possibly belonging
 to diffuse bands have been identified in spectra of Milky Way stars
 \citep{pwa86,destree09}. Of the few UV DIB candidates suggested, most
 lines lie too close to other metal absorption lines in the
 low-resolution composite spectrum to allow their identifications. Two
 exceptions are UV features at 1384.3~{\AA} and 1490.1~{\AA}, which
 are not detected in the composite afterglow spectrum to a 3\,$\sigma$
 limit of 60 m{\AA}.  The features observed in stellar spectra are
 even weaker than this detection limit, so if UV DIBs are present in
 GRB spectra an even better S/N ratio is necessary to be able to
 detect them.

\subsection{High-ionization Lines}
In almost all low redshift starburst galaxies, the
\ion{Si}{iv}\,$\lambda$1400 and \ion{C}{iv}\,$\lambda$1550 lines are
quite strong with rest-frame equivalent widths $>$3\,{\AA} as seen in
HST spectra of local starburst galaxies (C. Tremonti, private
communication). In the afterglow composite, the absorption lines are
weaker with $W_r$=1 and 2\,{\AA}, respectively. The transitions arise
as a combination of stellar wind features and intrinsic ISM
absorption. In starburst galaxies, the wind component is typically
recognized from their P-Cygni profiles. Contrary to direct spectra of
starbursts, the GRB afterglow spectra are not the direct emission
associated with massive stars, so with the absence of a recognizable
P-Cygni profile in the GRB afterglow spectra, a lower $W_r$ is
expected.

The \ion{C}{iv} and \ion{Si}{iv} lines typically extend over several
hundreds of km in GRB spectra, and have several components some of
which are saturated \citep[][; F09]{prochaska08b,fox08}. In the
low-resolution composite spectrum these lines have residual relative
fluxes $<0.7$, which suggests that they are composed of individual
saturated lines as explained in Sect.~\ref{sect:coldens}.

The highly ionized gas could potentially trace the medium surrounding
the GRB progenitors. Species like \ion{N}{v} and \ion{O}{vi} require a
stronger ionizing radiation field than that produced by O stars, but
which can be produced by a GRB event \citep{prochaska08}, although the
creation by collisional ionization in the ISM cannot be ruled out
\citep{fox08}. The \ion{N}{v} $\lambda\lambda$1238,1242 doublet is
distinguishable in some of the individual afterglow spectra at low
resolution, while in high resolution spectra reveal that these lines
are ubiquitous \citep{prochaska08,fox08}. These lines are not easily
recognizable in the composite spectrum, because they are blended with
the red wing of the \lya\ line, in particular in spectra with high
signals and very high \ion{H}{i} column densities. In the composite
spectrum the \ion{N}{v} lines appear as a slight deviation from the
broad wing profile of \lya. A more detailed analysis of these line
requires a higher spectral resolution.

\section{Comparison with starburst galaxy spectra}
\label{sect:lbg}
With a high S/N composite it is relevant to look for differences in
the absorption line properties compared to what is known about other
galaxies. With a spectral range covering from the UV to the rest-frame
optical wavelengths, a comparison with local starburst galaxies is
justified, but given the redshift difference it is more appropriate to
compare with starburst galaxies at higher redshifts. Such a comparison
is hampered by the faintness of the distant galaxies, but in a few
cases a good signal from high redshift galaxies have been obtained.

A low resolution composite spectrum from 811 Lyman break galaxies
(LBGs), which has a measured S/N of $\sim40$ and a rest-frame
wavelength coverage from 920 to 2000~{\AA} is presented in
\citet{shapley03}.  The composite LBG spectrum has been created by
averaging the spectra of galaxies at $z>3$ with magnitudes
$R\lesssim25.5$, and the composite spectrum has a dispersion of 1
{\AA} pixel$^{-1}$. With an average redshift $z\sim3$, this is
comparable to the GRB composite spectrum. Another example is the
magnified signal from high redshift galaxies lensed by foreground
galaxies or clusters. The prototype LBG spectrum obtained with
Keck/ESI of MS~1512-cB58 (cB58 for short) at $z=2.7276$
\citep{pettini00,pettini02}, which covers 1050--2800 {\AA} in the 
rest-frame and has an S/N$\sim$2--15, presents another valid comparison for
the GRB composite. Finally, we also compare the GRB composite with a
UV composite spectrum of UV bright galaxies at $1.3<z<2$ from the
Gemini Deep Deep survey (GDDS) \citep{savaglio04b}, which covers the
wavelength range 2000--3000~{\AA} with an S/N of about 15.

Fine-structure lines can arise in a medium which is being pumped by UV
radiation, or in very high density regions
\citep{silva02}. \ion{Fe}{ii}* fine-structure lines have been detected
in several GRB sight lines but not in quasar absorption systems
\citep{chen05}, and locally only in the dense wind from $\eta$ Carinae
\citep{gull95}. The UV pumping scenario and the time variations of the
absorption lines in time series observations of individual bursts
indicate a decrease of the ionizing radiation from the GRB
\citep{prochaska06b}. In this scenario, one would only expect the
presence of \ion{Fe}{ii}* fine-structure lines to be related to the
environment in the GRB host galaxy, and generally not to be present in
the ISM in normal starburst galaxies.

\begin{figure*} 
\figurenum{6} 
\plotone{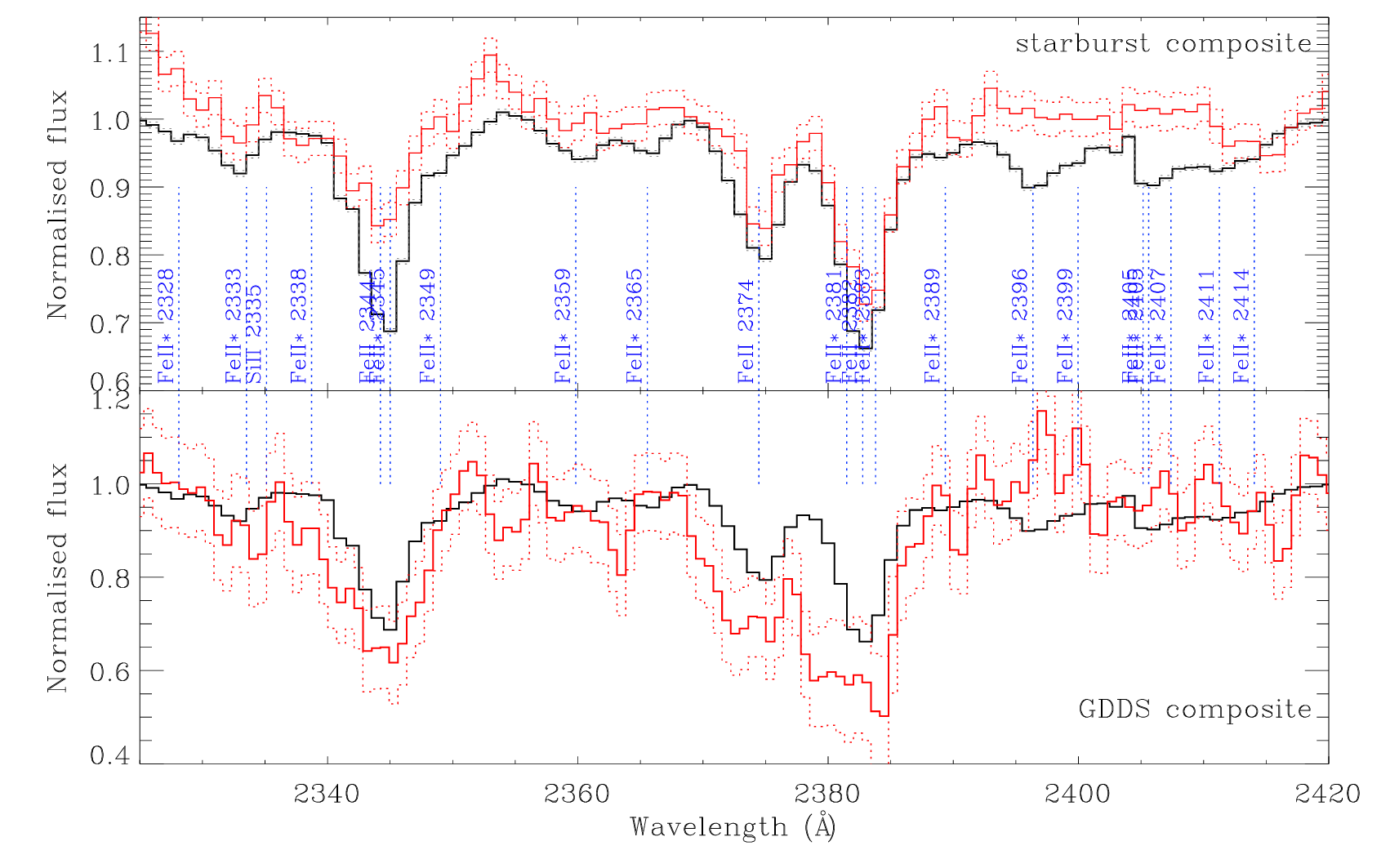} 
\caption{Section of the GRB afterglow composite spectrum (in black)
  compared to a starburst galaxy composite spectrum in red in the
  \textit{upper panel} (from C. Tremonti). The dotted lines correspond
  to 1$\sigma$ deviations from the composite. While strong
  \ion{Fe}{ii}* fine-structure transitions are present in the GRB
  composite, only the those from the ground state are present in the
  starburst composite. The \textit{lower panel} shows the same
    region for the GDDS composite spectrum.}
\label{fig:compare_plot}
\end{figure*}

Since long duration GRBs explode in star forming galaxies
\citep{christensen04b}, we here investigate if the fine-structure
lines could be present generally in starburst galaxies which have
particularly large SFRs, or strong emission in the UV,
or emission arising from dense environments.  In a composite spectrum
created from local starburst galaxies (C. Tremonti private communication)
most of the metal absorption lines in the range 2150--3250~{\AA}
identified in the GRB composite are also present (See
Figure~\ref{fig:compare_plot}). The strongest transition is that from
\ion{Fe}{ii}$^*$~$\lambda$2396 which is clearly detected in the GRB
composite, but due to the lower quality of the starburst composite
(S/N$\sim$30), we cannot claim a clear presence of \ion{Fe}{ii}* fine
structure lines. A better signal from a starburst galaxy is needed to
establish if these lines are only present in the most extreme
conditions in the region around a GRB explosion either by being pumped
by the hard UV radiation, or by having a specially high density.

In terms of tracing starburst galaxies, LBGs are brighter and more
massive than GRB hosts at the same redshifts due to the selection of
bright sources for the LBG samples, whereas GRBs trace star forming
galaxies irrespectively of their brightness. The same selection
effects play a role for the galaxies in the GDDS, while cB58 has an
intrinsic brightness of a typical $L$* galaxy. None of the comparison
spectra show \ion{Fe}{ii}* lines; in the case of cB58 the lines around
1600, 2300, and 2600~{\AA} could have been detected if they had been
there. The non-detection of fine-structure lines in starburst galaxies
is perhaps not surprising given that \ion{Fe}{ii}* lines in afterglow
spectra decay rapidly after the burst, and on that basis the lines are
not expected to be present on the much longer timescales present for
regions where massive stars are formed.

With a few exceptions, the $W_r$ measured for cB58 in
\citet{pettini02} have typical values about twice the strengths of
those in the GRB composite. The exceptions are the \ion{Fe}{ii}
$\lambda\lambda$1608,1611 blend, \ion{Cr}{ii},
\ion{Zn}{ii}~$\lambda$2026 and \ion{Cr}{ii},
\ion{Zn}{ii}~$\lambda$2062 blends, and the \ion{Mn}{ii} $\lambda$2577
line, all of which have roughly the same line strengths as in the GRB
composite. In contrast, the LBG and GDDS spectra have in general line
strengths in between the GRB composite and cB58 \citep[see also Table
  2 in][]{savaglio04b}.

LBGs and GDDS galaxies are relatively more massive, have higher SFRs,
and have stronger outflows driven by massive stars and supernovae
relative to the GRB hosts. The SFRs derived from the UV continuum flux
relative to the GRB hosts luminosities are comparatively high
\citep{christensen04b}. In contrast the specific SFR (SFR per unit
stellar mass) is found to be low compared to other high redshift
galaxies \citep{castro-ceron06,castro-ceron10}, while
\citet{savaglio09} find the contrary. It is therefore still an open
issue as to which degree the line strengths in GRB hosts and other
high redshift galaxies are related to their specific SFRs.

The \ion{C}{iv} $\lambda\lambda$1548,1550 lines are stronger in all
comparison spectra.  This line arises as a combination of ISM and
stellar, outflow driven, absorption components. Since the GRB hosts
are generally very faint, the emission from the host does not
dominate, such that no P-Cygni profiles are expected in the GRB
afterglow spectra. Other exceptions are the \ion{Si}{ii}* lines which
are found in absorption in the afterglow composite, while these lines
are found in emission in the cB58 spectrum and the LBG composite.

\section{Column densities}
\label{sect:coldens}
The hydrogen column density for cB58 is about one order of magnitude
below that in the GRB composite: log\,$N$(\ion{H}{i})~=~20.85
resulting in a higher metallicity of 40\% solar \citep{pettini02}.
Since the widths of the metal absorption lines in the GRB spectrum are
generally lower than in other high redshift galaxies, one is led to
the question of what is the average metallicity in GRB lines of sight.
However, this is a complex issue to address using the GRB composite
spectrum, because each absorption line does not measure the same
samples of individual GRB lines of sight.  Any composite spectrum will
carry along the scatter of properties of the individual objects
(galaxies or GRB afterglows) that went into it in the first place, and
a physical interpretation is hampered by this. Still, the general
features in a composite spectrum can be derived, as done from the LBG
composite with its interstellar absorption lines \citep{shapley03} and
for the GDDS composite, where the trends of metal column densities,
depletion patterns and extinction were derived for galaxies at
$1.4<z<2$ \citep{savaglio04b}. Compared to these studies, the GRB
afterglow composite may present further complications, because the
galaxies probed by the random sight line of a GRB through their
ISMs are not as uniformly selected.

Bearing this in mind we analyse the metal lines to derive column
densities for a wide range of elements. Since the spectra are obtained
at low spectral resolution, we use a standard single component
curve-of-growth (COG) method to calculate the column densities. For
GRB spectra which have absorption lines with multiple, saturated
components spread over a velocity range of several 100 km~s$^{-1}$,
the simple COG method underestimates the true metal column densities
and the metallicities can be larger by up to 1 dex in the worst case
scenario \citep{prochaska06c}.

In the case of \ion{H}{i}, its column density is determined instead
from Voigt profile fitting to the red wing of the absorption line.  In
the COG fit, the best fit Doppler parameter ($b$) is determined using
ten individual \ion{Fe}{ii} lines. Including all the uncertainties of
the lines, a best fit yields $b=44\pm2$ km~s$^{-1}$ and
log\,$N$(\ion{Fe}{ii})=15.62$\pm$0.05. However, as demonstrated in
\citet{prochaska06c}, the weaker lines in the linear part of the COG
have underpredicted strengths, and the true column density may be
larger than given by the best fit. Instead, we choose to tie the fit
to the four weaker lines (\ion{Fe}{ii} $\lambda$1611, $\lambda$1901,
$\lambda$2249, and $\lambda$2260) as illustrated in
Fig.~\ref{fig:cogplot}.  Since the widths of the stronger iron lines
have been calculated by deblending them from the \ion{Fe}{ii}*
fine-structure lines, some contamination may still occur, in
particular for the \ion{Fe}{ii} $\lambda$2382 and \ion{Fe}{ii}
$\lambda$2600 lines. The best fit is determined from $\chi^2$
minimization, which gives a Doppler parameter $b=34\pm2$~km~s$^{-1}$
and log\,$N$(\ion{Fe}{ii})~=~15.70$\pm$0.07. All \ion{Fe}{ii} lines
follow a single curve well both in the linear part, while the points
in the flat part show some scatter around the best fit COG. To assess
the amount of contamination by the fine-structure lines, we examine
the composite spectrum created from the late-time sample in
Sect.~\ref{sect:temporal}. We measure the $W_r$ for the 10
\ion{Fe}{ii} lines and find the best fit with a smaller Doppler
parameter: $b=20\pm2$~km~s$^{-1}$ and
log\,$N(\ion{Fe}{ii})$~=~15.60$\pm$0.10.

Since low resolution data indicate a higher $b$ parameter than when
observed at high resolution, because it measures the spread of the
lines \citep{prochaska06c} rather than the intrinsic velocity of the
gas, it is relevant to compare with individual measurements of $b$
from detailed high resolution spectra. Typical Doppler parameters
measured in various GRB afterglow spectra are between 7 and 15
km~s$^{-1}$ \citep{fiore05,chen07,thoene07,delia09,delia09b}, while
\citet{vreeswijk07} found $b=25$ km~s$^{-1}$ for GRB\,060418.  Hence,
the reason for our finding of a higher $b$ is partly due to blending
with iron finestructure lines, while a minor remaining part could be
caused by intrinsic line spreads of different components in a line.

To derive column densities for other elements, we use the best-fit $b$
parameter from the fit to the iron lines. Column densities for all the
elements are listed in Table~\ref{tab:coldens}. We avoid using lines
which are obviously blended with other elements as for example the
\ion{O}{i} $\lambda$1302 and \ion{Si}{ii} $\lambda$1304 blend.  To
derive the column density of \ion{Zn}{ii}, we subtract the
contribution of the \ion{Cr}{ii}\,$\lambda$2062 to the total width of
the line. From fit of the other two \ion{Cr}{ii} lines to the COG we
find that $W_r^{\ion{Cr}{ii}\lambda2062}=0.17$ {\AA}, whereas the
\ion{Cr}{ii}\,$\lambda$2026 line makes an insignificant contribution
to the \ion{Zn}{ii}\,$\lambda$2026 line. If the column density of
\ion{Mg}{i} is low as the best fit suggests in
Sect~\ref{sect:coldens}, then the contribution of \ion{Mg}{i}
$\lambda$2026 to the line blend is 0.05~{\AA}, which we subtract for
the fit to the \ion{Zn}{ii} column density.

We checked whether the composite spectrum is reliable for the analysis
of the metal column densities by comparing the fits with the spectrum
derived from Eq. (3). Using the ten \ion{Fe}{ii} lines, we get a
consistent result with the best fit of
log\,$N$(\ion{Fe}{ii})~=~15.90$\pm$0.10 and
$b=30\pm2$~km~s$^{-1}$. With a lower S/N, this composite does not
allow us to derive column densities of all the weak lines and rare
elements.

The elements are shown in Fig.~\ref{fig:cogplot} as different symbol
shapes. This plot illustrates that in general, the absorption lines
fit well to the \ion{Fe}{ii} COG apart from at low values of
log\,($Nf\lambda$~(cm$^{-1}$))$\lesssim$7, where the scatter may have a
physical explanation. To mention a few elements in particular, the
\ion{Ti}{ii} lines are relatively weak, and the scatter for the six
detected lines for this element in Fig.~\ref{fig:cogplot} is
substantial. Also some of the \ion{Mg}{i} lines are out-lying relative
to the COG. The main reason is possibly the sampling of the absorption
lines by different GRBs. Since the \ion{Mg}{i} lines lie in the range
1747--2852~{\AA} and \ion{Ti}{ii} lines in the range 1906--3384~{\AA},
each absorption line must sample a different set of GRBs. In these
cases, we exclude points with log\,($Nf\lambda)\lesssim$7,
but warn that in reality the column densities for these elements could
be $\sim$1~dex \textit{higher} than listed in Table~\ref{tab:coldens}.
The same tendency is observed for \ion{Fe}{i} lines.  At the other end
of the scale, \ion{Si}{ii} $\lambda$1260 and \ion{Si}{ii}
$\lambda$1808 are blended with their respective fine-structure lines,
and are not possible to deblend. As a consequence, the \ion{Si}{ii}
lines in Fig.~\ref{fig:cogplot} scatter and we base the fit for the
column density mostly on the isolated \ion{Si}{ii} $\lambda$1526
line. Since the only singly ionized carbon line (\ion{C}{ii}
$\lambda$1534) detected in the composite is blended with \ion{C}{ii}*
$\lambda$1335, we cannot calculate its column density
accurately. Instead, we assume that the two lines contribute roughly
equally to the total line strength such as observed in GRB\,050730
\citep{prochaska07b}.

The majority of the singly ionized species listed in
Table~\ref{tab:coldens} have ionization potentials above 13.6~eV, and
with the high column densities these are generally the dominant
species in DLAs. The metallicity of the different elements are given
relative to the solar value [X/H]~=~log\,($N_{\mathrm{X}}/N_{\mathrm{H}}$) --
log\,($N_{\mathrm{X}}/N_{\mathrm{H}}$)$_{\odot}$ in column 5 in
Table~\ref{tab:coldens}.  Where column densities from different
ionization levels are derived, the metallicity corresponds to their
sum. Different elements show relative metallicities in the range of
1\%--10\% solar. In high column density media, the dominant state of Ca
is \ion{Ca}{iii}, so the low relative metallicity of \ion{Ca}{ii} does
not reflect the real metallicity of this element. Also the
\ion{Mg}{ii} lines are saturated, so the metallicity is a lower limit.

\begin{figure*} 
\figurenum{7} 
\plotone{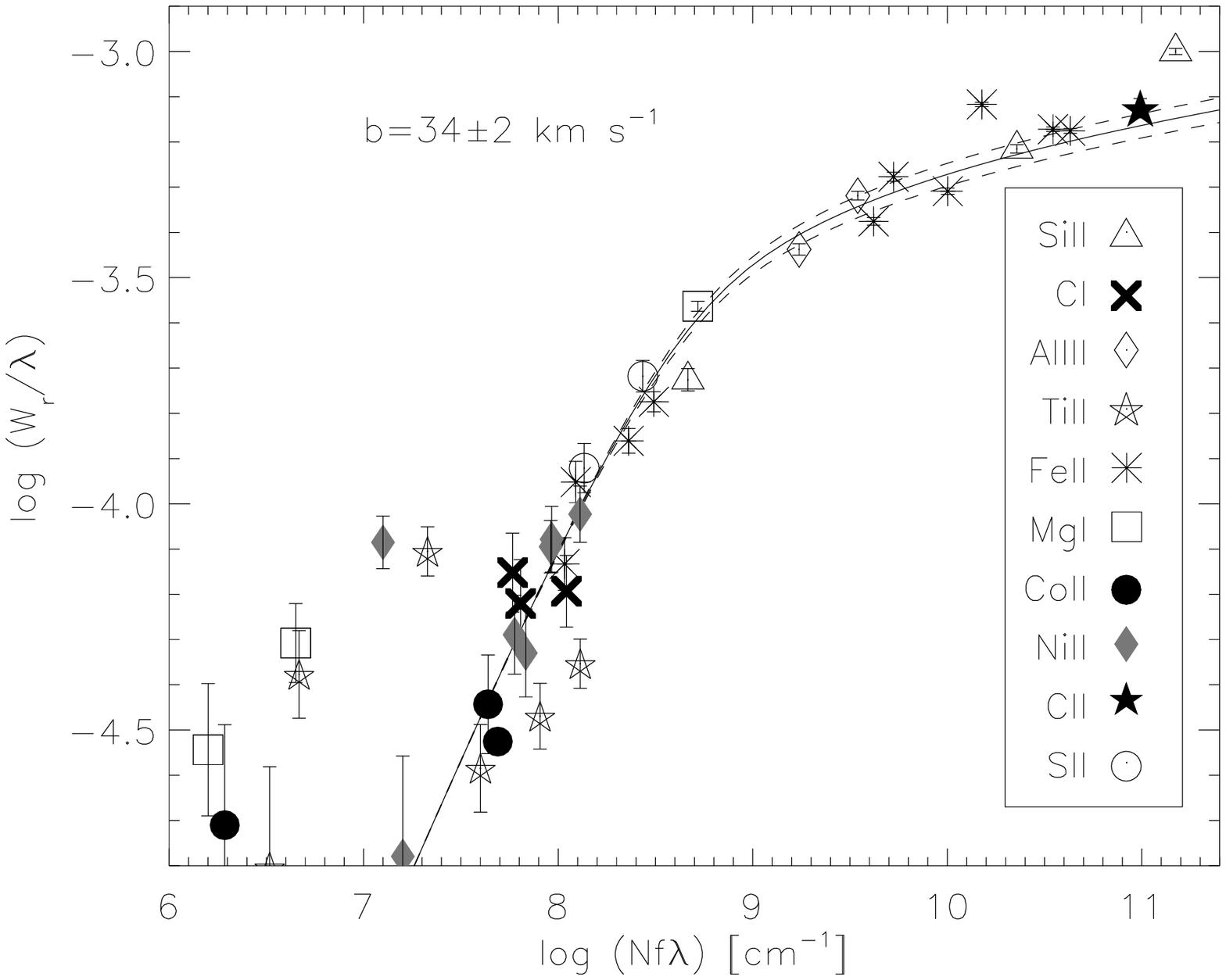} 
\caption{Curve-of-growth with the best Doppler parameter (and
  $\pm$1$\sigma$ uncertainties) fit estimated from ten \ion{Fe}{ii}
  lines. Other elements, are overplotted and all best fit column
  densities are listed in Table~\ref{tab:coldens}. }
\label{fig:cogplot}
\end{figure*}

\begin{figure*} 
\figurenum{7} 
\plotone{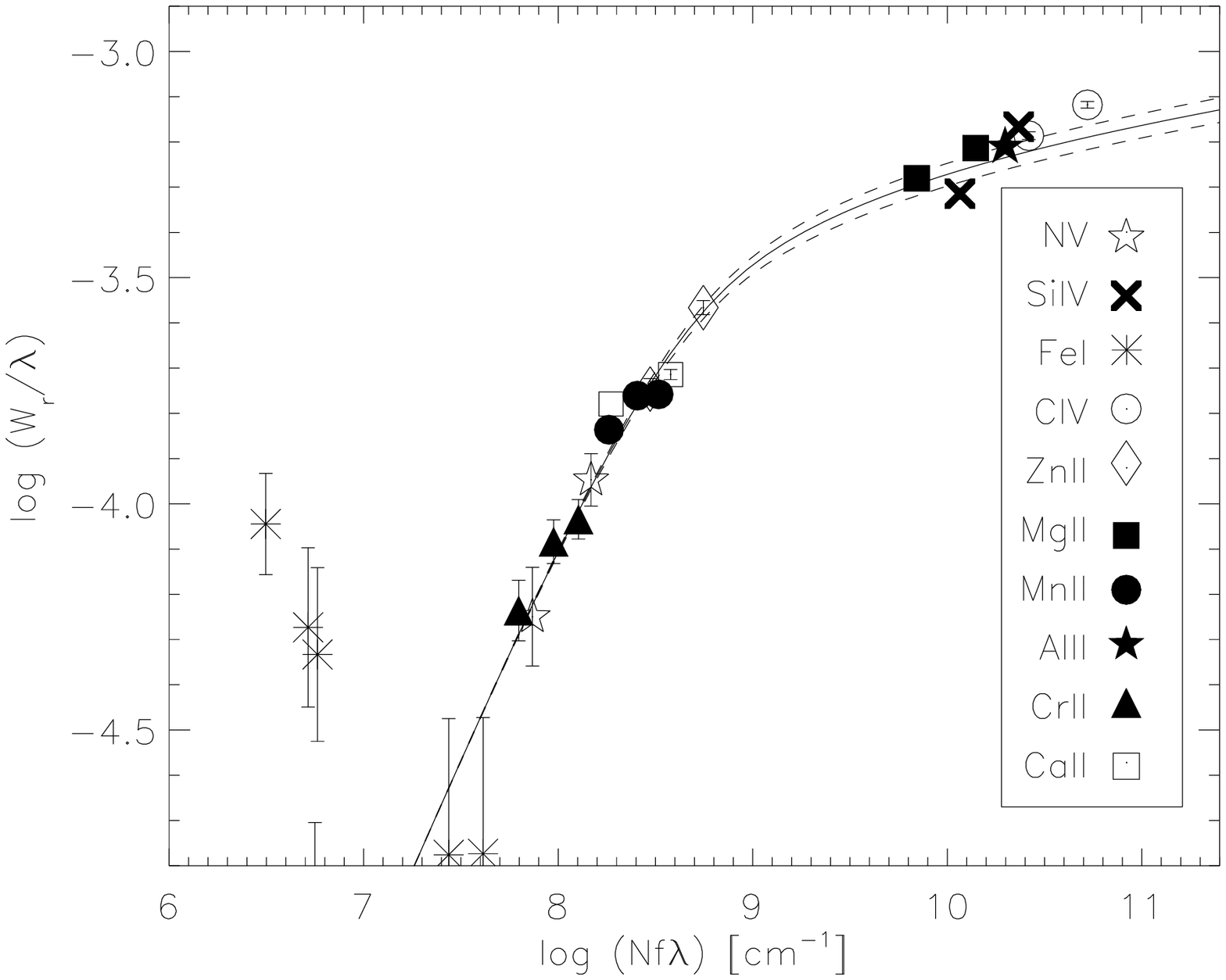} 
\vspace*{0.4cm}
\caption{\textit{continued}. Other elements are fit to the \ion{Fe}{ii} COG.}
\label{fig:cogplot}
\end{figure*}

To check the derived column densities for consistency, we also use the
apparent optical depth (AOD) method \citep{savage91}, which is valid
for non-saturated lines. In low resolution data, intrinsically
saturated lines are smoothed and do not appear saturated.  In practice
we find that lines with residual flux $\lesssim0.7$ in the composite
spectrum are most likely to be saturated, and using these lines will
only give a lower limit for the column densities. Column 6 in
Table~\ref{tab:coldens} lists the column densities of the various
species, where the individual lines were integrated from --600
km~s$^{-1}$ to +600 km~s$^{-1}$. We avoided as best possible the
individual saturated lines as well as lines which are blended with
other species. Comparing columns 2 and 6, the agreement between column
densities derived from the COG and the AOD methods is good. The
exceptions are \ion{Fe}{i} and \ion{Mg}{i}, where the AOD method
depends strongly on what lines are used to derive the column
densities. For example the \ion{Mg}{i} $\lambda$2852 suggests a much
lower column density log\,$N$=13.06$\pm$0.10 compared to that derived
from the other two lines (\ion{Mg}{i} $\lambda$1748 and \ion{Mg}{i}
$\lambda$1828 which indicate log\,$N$=14.20$\pm$0.10). The large
difference in column density can be due to the fact that the
\ion{Mg}{ii} $\lambda$2852 line is saturated.  Further, these lines
also exhibit a large scatter in the COG plot, and may as explained
above be the result of a physical difference in the medium probed by
the GRBs.

\section{Discussion and future prospects}
\label{sect:disc}
Having analyzed a series of parameters, which may affect the strengths
of the metal absorption lines in GRB afterglows we arrive at the
following results.
\begin{itemize}
\item Variations in excited states are apparent.
\item No variations in the ground states.
\item No correlation with GRB energy release.
\item No correlation with optical afterglow brightness one day after
  the GRB.
\item Dark bursts have stronger lines.
\end{itemize}
These results suggest that the medium surrounding the GRB at a
significant distance is mainly responsible for the strengths of the
absorption lines, and that the GRB itself plays a minor role.  Only
the excitation of the fine-structure lines by the immediate ionizing
radiation appear to reflect the impact of the GRB explosion. That some
afterglows are bright and have either strong or weak absorption lines,
may just reflect the diverse nature of the ISM along different lines
of sight through high redshift galaxies.

Characterizing the physical parameters from the composite is difficult
because each absorption line has a different sample of individual GRB
spectra contributing to it.  As such the composite is not relevant for
determining the physical conditions present around a single GRB
because it is a global average of different lines of sight at a range
of redshifts, where diverse conditions such as metallicities, abundance
ratios, and densities may prevail. The results are therefore not
directly applicable to a particular GRB environment. Biases may also
hamper the comparison to individual GRB afterglows observed in the
future, since the individual spectra which dominate the composite
spectrum are relatively bright, and are not affected severely by dust
obscuration.

These effects will inevitably influence the column densities of
species, where their individual absorption lines lie far apart in
wavelength. Although we estimate metallicities from several elements
to be 1\%--10\% solar, we do not attempt to analyze the abundance
patterns in comparison to the solar or depletion onto dust grains,
because of the inherent uncertainty of the GRB sample. Interestingly,
previous studies of individual GRB afterglows observed with low
resolution spectroscopy have indicated ISM metallicities from
10\%--100\% solar
\citep{savaglio06,fynbo06,vreeswijk06,berger06,prochaska07,delia07},
although a very low metallicity is reported for GRB\,090926
\citep{rau10}.  Conversely, afterglows observed at high spectral
resolution show relatively lower metallicities \citep[average 2\%
  solar in][]{ledoux09}. The composite low resolution spectrum
presented here is in better agreement with the high-resolution results
and the more metal poor systems observed at low resolution. For
multi-phased absorption systems the column densities and metallicities
can be underestimated using a COG analysis of low spectral resolution
data \citep{prochaska06c}, which would then cause an even stronger
discrepancy between the high- and low-resolution samples.

Although the composite spectrum cannot be used as an absolute
reference scale to measure the relative strengths of a set of
absorption lines, it can be used to discriminate special bursts which
truly have outlying properties such as the pure high-ionization lines
in GRB\,090426 \citep{levesque10,thoene10}. With the recent advent of
the wide wavelength coverage (3000--25\,000~{\AA}), second generation
VLT spectrograph, X-shooter \citep{dodorico06}, this composite
spectrum can serve as a reference for line identifications \citep[see
  e.g.,][]{deugarte10} in future GRB afterglow studies.
A similar composite spectrum made from higher spectral resolution data
would help to disentangle any of the temporal variations in the strong
blends of lines from fine-structure and non-excited transitions.

The composite spectrum can also be used to derive the redshifts of
very noisy spectra using a cross-correlation technique. We attempted
this for GRB\,060614 which did not show any detectable absorption
lines in its spectrum (F09), but fail to determine a reliable redshift
probably because of the absence of absorption lines rather than a bad
S/N ratio. On the other hand, we determine tentative redshifts for
GRB\,070129 ($z=2.985$), GRB\,070508 ($z=2.790$), and GRB\,080523
($z=2.050$) consistent with possible, albeit noisy detections of
either \ion{Si}{iv}, \ion{C}{iv} or \ion{Al}{iii} lines in their
spectra, and also consistent with the upper redshift limits in F09.

An important future study is to investigate the effect of the location
of the GRB with respect to the host galaxy. The locations of high
redshift GRBs follow the light distributions, and are typically found
close to the brightest part of their host galaxy
\citep{bloom02a,fruchter06,svensson10}. We do not have the information
of the spatial offsets available to make a comparison here. Logically,
one could expect stronger absorption lines arising from GRBs which are
located at smaller impact parameters to their hosts, but this
hypothesis has to be verified first. Another line of investigation
would be the correlation of absorption line strength versus the host
galaxy brightness. Again we do not have sufficient data available to
make this investigation. As discussed in Sect.~\ref{sect:lbg}, LBG
galaxies show generally stronger lines, but are also selected among
the intrinsically brighter high redshift galaxies which presumably are
more massive and metal-rich than the average GRB host. Combined with
the brightness, it would also be an interesting extension to
investigate the correlations of absorption line strengths with the
SFRs and specific SFRs of the GRB hosts.

\acknowledgments 

We thank the anonymous referee for the constructive comments.  The Dark
Cosmology Centre is funded by the Danish National Research Foundation.
We thank Hsiao-Wen Chen, Sandra Savaglio, Palle M{\o}ller, Max
Pettini, Sergio Campana, Daniele Malesani and Cedric Ledoux for
helpful discussions. We also thank Christy Tremonti for supplying the
starburst galaxy composite spectrum, Sandra Savaglio for the GDDS
composite, Chuck Steidel for the cB58 spectrum, and Josh Bloom and
Hsiao-Wen Chen for sharing the reduced LRIS and GMOS spectra, and
finally Paul Vreeswijk for sharing the FORS data.


\bibliographystyle{apj}
\bibliography{ms_lc}
\clearpage

\begin{deluxetable}{lcc}
\tablecaption{Weighted Composite Afterglow Spectrum}
\tablewidth{0pt}
\tablehead{
\colhead{$\lambda$ ({\AA})} & \colhead{$f_{\lambda}$} &
\colhead{$f_{\lambda}$ uncertainty} 
}
\startdata
701 &  0.003 & 0.027 \\
702 &  0.018 & 0.027 \\
703 &  0.000 & 0.027 \\
704 &  0.000 & 0.027 \\
705 &  0.000 & 0.027 \\
706 &  0.026 & 0.026 \\
707 &  0.000 & 0.026 \\
708 &  0.001 & 0.025 \\
709 &  0.001 & 0.025 \\
710 &  0.000 & 0.025 \\
711 &  0.000 & 0.025 \\
712 &  0.007 & 0.024 \\
713 &  0.030 & 0.024 \\
714 &  0.012 & 0.024 \\
\enddata
\label{tab:composite_spec}
\tablenotetext{}{Relative flux in the composite spectrum and its
  associated error spectrum.  This table is available in its entirety
  in a machine-readable form in the online journal. A portion is shown
  here for guidance regarding its form and content.}
\end{deluxetable}

\vspace*{0.5cm}

\LongTables

\begin{deluxetable}{lllll}
\tablecaption{Absorption Line List}
\tablewidth{0pt}
\tablehead{
\colhead{Line ID} & \colhead{$\lambda_{\mathrm{lab}}$
  ({\AA})\tablenotemark{c}} & \colhead{$\lambda_{\mathrm{obs}}$} &
\colhead{$W_r$ ({\AA})} &  \colhead{notes}}

\startdata
\ion{H}{i}    & 1215.67 &  & 73   \\
\ion{N}{v}    & 1238.82 & 1239.01 & 0.14$\pm$0.02 \\
\ion{N}{v}    & 1242.80 & 1243.01 & 0.07$\pm$0.02 \\
\ion{S}{ii}   & 1250.58 & 1249.95 & 0.15$\pm$0.02  \\
\ion{S}{ii}   & 1253.52 & 1252.91 & 0.24$\pm$0.02  \\

\ion{S}{ii}   & 1259.52 & \multirow{2}{*}{1259.95} & \multirow{2}{*}{1.26$\pm$0.02} &  a \\
\ion{Si}{ii}  & 1260.42 &         &      & a\\
\ion{Si}{ii}* & 1264.74 & 1264.58 & 0.66$\pm$0.02  \\
\smallskip\\
\ion{C}{i}    & 1277.25 & \multirow{2}{*}{1277.51} & \multirow{2}{*}{0.09$\pm$0.02} &  \\   
\ion{C}{i}*   & 1277.28--1280.58  & &  &  blends \\   
\smallskip\\
\ion{O}{i}    & 1302.17 & \multirow{3}{*}{1302.97} & \multirow{3}{*}{2.29$\pm$0.02} &  a\\      
\ion{Si}{ii}  & 1304.37 & &  & a\\   
\ion{O}{i}*   & 1304.86 & &  & a\\
\smallskip\\
\ion{Si}{ii}* & 1309.28 & 1309.32 & 0.27$\pm$0.02 &  \\   
\ion{Ni}{ii}  & 1317.22 & 1318.00 & 0.11$\pm$0.02 &   \\   
\smallskip\\
\ion{C}{i}    & 1328.83 & \multirow{2}{*}{1328.95}& \multirow{2}{*}{0.08$\pm$0.01} &  a\\   
\ion{C}{i}*,**& 1329.09--1329.60 & & &  blends\\   
\smallskip\\
\ion{C}{ii}   & 1334.53 & \multirow{2}{*}{1334.99} & \multirow{2}{*}{1.73$\pm$0.02} &  a\\   
\ion{C}{ii}*  & 1335.71 & & &  a\\   
\smallskip\\
\ion{Cl}{i}   & 1347.24 & 1347.01 & 0.20$\pm$0.02 &  b \\   
\ion{Ni}{ii}  & 1370.13 & 1370.51 & 0.13$\pm$0.02 &   \\   
\ion{Si}{iv}  & 1393.76 & 1393.51 & 0.95$\pm$0.02 &  \\   
\ion{Si}{iv}  & 1402.77 & 1402.61 & 0.68$\pm$0.02 &  \\   
\smallskip\\
\ion{Ga}{ii}  & 1414.40 & \multirow{2}{*}{1414.95} & \multirow{2}{*}{0.16$\pm$0.02} &  a, b \\   
\ion{Ni}{ii}  & 1415.72 & & &  a, b  \\   
\smallskip\\
\ion{CO}      & 1419.0  & 1419.00 & 0.06$\pm$0.01 &  b  \\   
\smallskip\\
\ion{Ni}{ii}  & 1454.84 & \multirow{2}{*}{1456.00} & \multirow{2}{*}{0.08$\pm$0.02} &  \multirow{2}{*}{a, b} \\
\ion{Zn}{i}   & 1457.57 & & &   \\   
\smallskip\\
\ion{CO}      & 1477.5  & 1478.4  & 0.21$\pm$0.02 &    \\   
\ion{Si}{ii}  & 1526.71 & 1526.02 & 0.93$\pm$0.02 &   \\   
\ion{Si}{ii}* & 1533.43 & 1533.51 & 0.42$\pm$0.02 &  \\  
\ion{Co}{ii}  & 1539.47 & 1539.00 & 0.03$\pm$0.01 &  b \\   
\smallskip\\
\ion{C}{iv}   & 1548.20 & \multirow{2}{*}{1548.33} & \multirow{2}{*}{2.18$\pm$0.03} &   a \\   
\ion{C}{iv}   & 1550.77 &  & &   a \\   
\smallskip\\
\ion{C}{i}    & 1560.31 &  1559.51 & 0.09$\pm$0.02  & b\\
\ion{Fe}{ii}* & 1570.25 &  1569.92 & 0.08$\pm$0.02  & b\\
\ion{Fe}{ii}* & 1602.49 &  1602.56 & 0.08$\pm$0.01  & \\
\ion{Fe}{ii}  & 1608.45 &  1608.08 & 0.85$\pm$0.02  & a\\
\ion{Fe}{ii}  & 1611.20 &  1610.81 & 0.18$\pm$0.02  & a\\
\smallskip\\
\ion{Fe}{ii}* & 1618.47 &  1618.02 & 0.03$\pm$0.01 &  a\\
\ion{Fe}{ii}* & 1621.69 &  1621.60 & 0.11$\pm$0.02 &  a\\
\smallskip\\
\ion{Fe}{ii}* & 1629.16 & \multirow{2}{*}{1629.50} & \multirow{2}{*}{0.10$\pm$0.02} & \multirow{2}{*}{a}\\
\ion{Fe}{ii}* & 1631.13 & &  & \\
\smallskip\\
\ion{Fe}{ii}* & 1634.35 & \multirow{3}{*}{1637.55} & \multirow{3}{*}{0.50$\pm$0.02} & a \\
\ion{Fe}{ii}* & 1636.33 & &  & \\
\ion{Fe}{ii}* & 1639.40 &  & & \\
\smallskip\\
\ion{C}{i}    & 1656.93 & \multirow{2}{*}{1657.02} & \multirow{2}{*}{0.14$\pm$0.02} & a\\
\ion{C}{i}*,** & 1656.27--1658.12 & &  &  blends \\
\smallskip\\
\ion{Al}{ii}  & 1670.79 & 1670.58 & 1.04$\pm$0.02 &  \\
\ion{Si}{i}   & 1693.29 & 1693.03 & 0.07$\pm$0.02 &  \\ 
\ion{Ni}{ii}  & 1703.41  & 1702.48 & 0.14$\pm$0.02 &  \\
\ion{Ni}{ii}  & 1709.60 & 1709.45 & 0.08$\pm$0.02 &  \\
\ion{Ni}{ii}  & 1741.55 & 1742.02 & 0.14$\pm$0.02 & \\
\ion{Mg}{i}   & 1747.79 & 1748.02 & 0.05$\pm$0.01 &  b\\
\ion{Ni}{ii}  & 1751.92 & 1752.06 & 0.09$\pm$0.02 &  \\
\smallskip\\
\ion{Ni}{ii}  & 1804.47 & 1805.31 & 0.03$\pm$0.01 & a \\
\ion{S}{i}    & 1807.31 & \multirow{2}{*}{1808.14} & \multirow{2}{*}{0.29$\pm$0.02} & a\\ 
\ion{Si}{ii}  & 1808.01 & & & a\\
\smallskip\\
\ion{Si}{ii}* & 1816.93 & \multirow{2}{*}{1816.95} & \multirow{2}{*}{0.12$\pm$0.01} & a, b\\
\ion{Si}{ii}* & 1817.45 & & & a, b\\
\smallskip\\
\ion{Mg}{i}   & 1827.93 & 1827.87 & 0.09$\pm$0.02 &   \\
\ion{Ni}{ii}  & 1842.89 & 1844.27 & 0.12$\pm$0.02 &  b \\
\ion{Al}{iii} & 1854.72 & 1854.83 & 0.89$\pm$0.02 &  \\
\ion{Al}{iii} & 1862.79 & 1863.56 & 0.68$\pm$0.02 &  \\
\ion{Fe}{i}   & 1875.16 & 1874.46 & 0.10$\pm$0.01 &  a  \\
\ion{Fe}{i}   & 1883.78 & 1882.67 & 0.17$\pm$0.02 &  a  \\
\ion{Si}{iii} & 1892.03 & 1890.58 & 0.10$\pm$0.02  & b \\
\smallskip\\
\ion{Fe}{ii}  & 1901.77 & 1901.36 & 0.14$\pm$0.02 &  a \\
\smallskip\\
\ion{Ti}{ii}  & 1905.77 & \multirow{2}{*}{1905.99} & \multirow{2}{*}{0.11$\pm$0.02} & \multirow{2}{*}{a} \\
\ion{Ti}{ii}  & 1906.24 &   \\
\ion{Ti}{ii}  & 1910.75 & 1910.56 & 0.14$\pm$0.02 & a \\
\smallskip\\
\ion{Fe}{i}   & 1937.27 & 1938.04 & 0.09$\pm$0.02 & a  \\
\ion{Co}{ii}  & 1941.29 & 1941.97 & 0.07$\pm$0.01 & a, b\\
\ion{Co}{ii}  & 2012.17  & 2010.99 & 0.06$\pm$0.02 &  b\\
\ion{Cr}{ii}  & 2017.57  & 2017.61 & 0.08$\pm$0.02 &  b\\
\smallskip\\
\ion{Zn}{ii}  & 2026.14  & \multirow{3}{*}{2025.97} & \multirow{3}{*}{0.60$\pm$0.02} &  \multirow{3}{*}{a}\\
\ion{Cr}{ii}  & 2026.27  & \\
\ion{Mg}{i}   & 2026.48  & \\
\smallskip\\
\ion{Cr}{ii}  & 2056.26  & 2055.51 & 0.19$\pm$0.02 &  \\
\smallskip\\
\ion{Cr}{ii}  & 2062.23  & \multirow{2}{*}{2062.67} & \multirow{2}{*}{0.53$\pm$0.02} &  a\\
\ion{Zn}{ii}  & 2062.66  & &  &  a\\
\smallskip\\
\ion{Cr}{ii}  & 2066.16  & 2066.29 & 0.12$\pm$0.02 & a \\
\smallskip\\
\ion{Ni}{ii*} & 2166.23  & \multirow{2}{*}{2167.65} &\multirow{2}{*}{0.26$\pm$0.02} & \multirow{2}{*}{a} \\ 
\ion{Fe}{i}   & 2167.45  &  &  &  \\ 
\smallskip\\
\ion{Ni}{ii*} & 2175.22  & 2175.83 & 0.07$\pm$0.01 &  \\ 
\ion{Mn}{i}   & 2185.59  & 2186.96 & 0.51$\pm$0.02 &  \\ 
\ion{Ni}{ii}* & 2217.2   & 2217.02 & 0.24$\pm$0.01 &  \\ 
\ion{Ni}{ii}  & 2223.0   & 2224.98 & 0.07$\pm$0.01 &  \\ 
\ion{Fe}{ii}  & 2249.88 & 2250.02 & 0.31$\pm$0.02 &  \\ 
\ion{Fe}{ii}  & 2260.78 & 2261.00 & 0.38$\pm$0.02 &  \\ 
\ion{Ni}{ii}  & 2316.7  & 2316.36 & 0.19$\pm$0.02 &  \\ 
\smallskip\\
\ion{Fe}{ii}**& 2328.11  & 2327.65 & 0.10$\pm$0.02 & \\
\ion{Fe}{ii}* & 2333.52  & 2333.30 & 0.34$\pm$0.02 & \\
\smallskip\\
\ion{Fe}{ii}  & 2344.21  & \multirow{2}{*}{2346.01} & \multirow{2}{*}{1.74$\pm$0.02} & \multirow{2}{*}{a}\\
\ion{Fe}{ii}* & 2345.00  \\
\ion{Fe}{ii}* & 2349.02  & 2349.60 & 0.26$\pm$0.02 & a\\
\smallskip\\
\ion{Fe}{ii}* & 2359.83  & 2360.01 & 0.28$\pm$0.01 & \\
\ion{Fe}{ii}* & 2365.55  & 2365.40 & 0.18$\pm$0.01 & \\
\ion{Fe}{ii}  & 2374.46  & 2374.46 & 1.00$\pm$0.02 &  \\
\smallskip\\
\ion{Fe}{ii}* & 2381.49  & \multirow{3}{*}{2383.59}&\multirow{3}{*}{1.65$\pm$0.02} &  a\\
\ion{Fe}{ii}  & 2382.77 & &  &  a\\
\ion{Fe}{ii}* & 2383.79 & &  &  a\\
\smallskip\\
\ion{Fe}{ii}* & 2389.36 & 2389.40 & 0.18$\pm$0.02 &  a\\
\smallskip\\
\ion{Fe}{ii}* & 2396.15 & \multirow{3}{*}{2396.50} & \multirow{3}{*}{0.71$\pm$0.02} &  a\\
\ion{Fe}{ii}* & 2396.36 & &  & a\\
\ion{Fe}{ii}* & 2399.98 & &  & a\\
\smallskip\\
\ion{Fe}{ii}* & 2405.16 &\multirow{2}{*}{2406.04} & \multirow{2}{*}{0.40$\pm$0.02} & a\\
\ion{Fe}{ii}* & 2407.39 & &  & a\\
\smallskip\\
\ion{Fe}{ii}* & 2411.25 & \multirow{3}{*}{2411.68} & \multirow{3}{*}{0.51$\pm$0.03} & a\\
\ion{Fe}{ii}* & 2411.80 & &  & a\\
\ion{Fe}{ii}* & 2414.05 & &  & a\\
\smallskip\\
\ion{Sc}{ii}  & 2561.00 & \multirow{2}{*}{2562.52} & \multirow{2}{*}{0.18$\pm$0.01} & a\\
\ion{Sc}{ii}  & 2563.97 & & & a\\
\smallskip\\
\ion{Mn}{ii}  & 2576.88 & 2577.04 & 0.45$\pm$0.02 &  \\
\ion{Fe}{ii}  & 2586.65 & 2586.49 & 1.33$\pm$0.02 & a \\
\smallskip\\
\ion{Mn}{ii}  & 2594.50 & 2594.36 & 0.45$\pm$0.02 & \\
\ion{Fe}{ii}* & 2599.15 & \multirow{2}{*}{2600.23} & \multirow{2}{*}{1.85$\pm$0.03} &  a\\
\ion{Fe}{ii}  & 2600.17 & &  &  a\\
\smallskip\\
\ion{Mn}{ii}  & 2606.46 & \multirow{2}{*}{2607.53} & \multirow{2}{*}{0.56$\pm$0.01} &  a\\
\ion{Fe}{ii}* & 2607.87 & &  &  a\\
\smallskip\\
\ion{Fe}{ii}* & 2612.65 & \multirow{2}{*}{2613.11} & \multirow{2}{*}{0.51$\pm$0.01} &  a\\
\ion{Fe}{ii}* & 2614.61 & &  &   \\
\smallskip\\
\ion{Fe}{ii}* & 2618.40 & 2618.51 & 0.06$\pm$0.01   &  a\\
\smallskip\\
\ion{Fe}{ii}* & 2621.19 & \multirow{6}{*}{2629.82} & \multirow{6}{*}{0.90$\pm$0.02} & a\\ 
\ion{Fe}{ii}* & 2622.45 & &  & a\\
\ion{Fe}{ii}* & 2626.45 & &  & a\\
\ion{Fe}{ii}* & 2629.08 & &  & a\\
\ion{Fe}{ii}* & 2631.83 & &  & a\\
\ion{Fe}{ii}* & 2632.11  & &  & a\\
\smallskip\\
\ion{Fe}{ii}* & 2740.4   & 2739.45 & 0.07$\pm$0.01 &  b\\
\ion{Fe}{ii}* & 2747.9   & 2749.50 & 0.16$\pm$0.01 &  b\\
\ion{Fe}{ii}* & 2756.28   & 2756.50 & 0.08$\pm$0.01 &  b\\
\ion{Mg}{ii}   & 2796.35   & 2796.21 & 1.71$\pm$0.02 &  \\
\ion{Mg}{ii}   & 2803.53   & 2803.50 & 1.47$\pm$0.02 &  \\
\ion{Mg}{i}    & 2852.96   & 2852.97 & 0.78$\pm$0.01 &  \\
\ion{Fe}{i}    & 2967.77   & 2966.24 & 0.05$\pm$0.01 &  b \\
\ion{Fe}{i}    & 2984.44   & 2983.49 & 0.05$\pm$0.01 &  \\
\ion{Ti}{ii}   & 3073.88   & 3076.01 & 0.08$\pm$0.01 &  \\ 
\ion{Ti}{ii}   & 3242.93   & 3239.56 & 0.11$\pm$0.01 &  \\ 
\ion{Ti}{ii}   & 3384.74   & 3385.85 & 0.15$\pm$0.02 &  b \\
\ion{Ca}{ii}   & 3934.78   & 3933.97 & 0.76$\pm$0.02 &  \\
\ion{Ca}{ii}   & 3969.59   & 3969.98 & 0.66$\pm$0.02 &  \\
\ion{Ca}{i}    & 4227.92   & 4226.93 & 0.11$\pm$0.02 &  b\\
\ion{Mg}{H}    & 5209.45   & 5210.75 & 0.09$\pm$0.02 &  b
\enddata
\label{tab:abslist}
\tablenotetext{a}{Lines are blended.}
\tablenotetext{b}{Line only present in one of the two subsamples,
  when splitting the full sample by signal-to-noise.}
\tablenotetext{c}{Vacuum wavelengths.}
\end{deluxetable}

\vspace*{0.5cm}


\begin{deluxetable}{llllll}
\tablecaption{Absorption line list blueward of Ly$\alpha$ \label{tab:bluelist}}
\tablewidth{0pt}
\tablehead{
\colhead{Line ID} & \colhead{$\lambda_{\mathrm{lab}}$ ({\AA})} & \colhead{$\lambda_{\mathrm{obs}}$ ({\AA})}
 &  \colhead{notes}\\ }
\startdata

\ion{Si}{ii}   & 1206.500             &  1206.99 \\ 
\ion{N}{i}     & 1199.549--1200.710   & \multirow{2}{*}{1200.47} & \multirow{2}{*}{blend}\\          
\ion{Mn}{ii}   & 1199.391             & \\
\ion{Si}{ii}   & 1190.4158, 1193.2897 & \multirow{2}{*}{1192.96} & \multirow{2}{*}{blend} \\
\bigskip

\ion{C}{i}*,** & 1190.254--1194.686   & \\
\ion{C}{i}*,** & 1157.910--1160.51    & 1160.52  & a \\
\ion{Fe}{ii}   & 1142.3656--1144.9379 & 1143.00  & a \\
\ion{C}{i}*,** & 1138.384--1141.678   & 1138.20  & a \\
\ion{Fe}{ii}   & 1125.4478            & 1127.01  & \\
\bigskip

\ion{C}{i}*,** & 1121.453  - 1123.460 & 1121.49  & a \\
\ion{Fe}{ii}   & 1081.8748            & \multirow{2}{*}{1085.04} \\ 
\ion{N}{ii}    & 1083.990             &  \\
\ion{Ar}{i}    & 1066.660             & \multirow{3}{*}{1066.00}\\
\ion{Fe}{ii}   & 1062.152--1063.9718  &  \\
\ion{S}{iv}    & 1062.662             &  \\
\bigskip
\ion{Ar}{i}    & 1048.2199            & 1049.58 & \\
\ion{O}{vi}    & 1037.6167            & \multirow{2}{*}{1037.55}\\
\ion{C}{ii}    & 1036.3367            & \\
\ion{O}{vi}    & 1031.9261            & 1033.00 \\
\ion{H}{i}     & 1025.7225 (Ly$\beta$)& \multirow{2}{*}{1024.39} & a\\ 
\ion{Mg}{ii}   & 1025.9681,1026.1134  &  \\
\ion{Si}{ii}   & 1020.6989            &  \\
\bigskip
\ion{Si}{ii}   & 1012.502             & 1013.04 \\
\ion{N}{ii}    & 989.799              & \multirow{2}{*}{989.49} \\ 
\ion{Si}{ii}   & 989.8731             & \\
\ion{C}{iii}   & 977.020              & 979.08 \\
\ion{H}{i}     & 972.5368             & 973.03 \\
\bigskip
\ion{N}{i}     & 953.415,953.655      & 954.49 \\
\ion{H}{i}     & 949.7431             & 950.52 \\
\ion{H}{i}     & 937.8035             & 937.02 \\
\ion{H}{i}     & 930.7483             & 930.99 
\enddata
\tablenotetext{a}{Lines are blended.}
\end{deluxetable}

\pagebreak

\begin{deluxetable}{lcccc}
\tablecaption{GRB spectra contributing at Lyman limit}
\tablewidth{0pt}
\tablehead{
\colhead{GRB name} & \colhead{redshift} & \colhead{$\Delta\lambda_r$
  ({\AA})\tablenotemark{a}} & \colhead{S/N ($\lambda<912.(1+z)$)} & 
  \colhead{log\,$N$(\ion{H}{i} cm$^{-2}$)\tablenotemark{b}}
 }
\startdata
GRB\,050730  & 3.9692 & 800--900 & 0.44 & 22.10$\pm$0.10\\
GRB\,050908  & 3.3467 & 840--900 & 1.00 & 17.60$\pm$0.10\\
GRB\,060206  & 4.0559 & 800--900 & 0.17 & 20.85$\pm$0.10\\
GRB\,060210  & 3.9133 & 800--900 & 0.02 & 21.55$\pm$0.15\\
GRB\,060526  & 3.2213 & 845--900 & 0.12 & 20.00$\pm$0.15\\
GRB\,060707  & 3.4240 & 810--900 & 0.08 & 21.00$\pm$0.20\\
GRB\,060906  & 3.6856 & 878--900 & 0.10 & 21.85$\pm$0.10\\
GRB\,060927  & 5.4635 & 800--900 & 0.10 & 22.50$\pm$0.15\\
GRB\,061110B & 3.4344 & 873--900 & 0.52 & 22.35$\pm$0.10\\
GRB\,080603  & 2.6892 & 870--900 & 0.12 & 21.85$\pm$0.05\\
GRB\,080607  & 3.0368 & 800--900 & 0.04 & 22.70$\pm$0.15
\enddata
\label{tab:lylimit}
\tablenotetext{a}{Wavelength range to which each
  spectrum contributes to the composite blueward of the Lyman limit.}
\tablenotetext{b}{\ion{H}{i} column densities are adopted from F09.} 
\end{deluxetable}

\vspace*{0.5cm}

\begin{deluxetable}{llllll}
\tablecaption{Column Densities \label{tab:coldens}}
\tablewidth{0pt}
\tablehead{
\colhead{Element} & \colhead{log\,$N$(cm$^{-2}$) (COG)} &
\colhead{log\,(X/H)}&
\colhead{log\,(X/H)$_{\odot}$\tablenotemark{a}} & \colhead{[X/H]} &
\colhead{log\,$N$(cm$^{-2}$)  (AOD)}  }
\startdata
\ion{H}{i}    & 22.00$\pm$0.10 &                 &  \\

\ion{C}{i}    & 13.73$\pm$0.15 & --8.27$\pm$0.18 & --3.57 &   & 13.72$\pm$0.09\\
\ion{C}{ii}   & 16.76$\pm$0.12\tablenotemark{b}  & --5.24$\pm$0.16 &  &
--1.55\tablenotemark{c}   & $>$15.0 \\
\ion{C}{iv}   & 16.25$\pm$0.15 & --5.75$\pm$0.18 &        &   & $>$15.0\\ 
\ion{N}{v}    & 13.88$\pm$0.05 & --8.12$\pm$0.11 & --4.17 &   & 13.90$\pm$0.10\\
\ion{Mg}{ii}  & 14.91$\pm$0.06 & --7.09$\pm$0.12 & --4.40 &
$>$--2.68\tablenotemark{c} &  $>$14.0 \\
\ion{Mg}{i}   & 13.06$\pm$0.10 & --8.94$\pm$0.14 &  &  & 14.20$\pm$0.15\\
\ion{Al}{ii}  & 14.80$\pm$0.20\tablenotemark{d}  & --7.20$\pm$0.22 &
--5.55 & --1.46\tablenotemark{c} & $>$13.4 \\
\ion{Al}{iii} & 14.54$\pm$0.20 & --7.46$\pm$0.22 &   &   &  $>$14.0 \\
\ion{Si}{ii}  & 16.07$\pm$0.22\tablenotemark{e} & --5.93$\pm$0.24  &
--4.49 & --1.27\tablenotemark{c} & 15.74$\pm$0.06\\
\ion{Si}{iv}  & 15.50$\pm$0.20 & --6.50$\pm$0.22 &        &        & $>$14.1 \\
\ion{S}{ii}   & 15.37$\pm$0.05 & --6.63$\pm$0.11 & --4.88 & --1.75 & 15.35$\pm$0.12  \\
\ion{Ca}{ii}  & 13.17$\pm$0.12\tablenotemark{f} & --8.83$\pm$0.16 & --5.66 & $>$--3.17 & 13.05$\pm$0.15 \\
\ion{Ti}{ii}  & 13.03$\pm$0.25 & --8.97$\pm$0.27 & --7.05 & --1.92 & 12.77$\pm$0.23 \\
\ion{Cr}{ii}  & 13.77$\pm$0.05 & --8.23$\pm$0.11 & --6.36 & --1.87 & 13.68$\pm$0.09 \\
\ion{Mn}{ii}  & 13.56$\pm$0.08 & --8.44$\pm$0.13 & --6.57 & --1.87 & 13.37$\pm$0.03 \\
\ion{Fe}{ii}  & 15.70$\pm$0.07 & --6.30$\pm$0.12 & --4.50 &
--1.80\tablenotemark{c}  & 15.49$\pm$0.10\\ 
\ion{Fe}{i}   & 13.50$\pm$0.15\tablenotemark{g} & --8.50$\pm$0.18  &
& &  14.90$\pm$0.10 \\
\ion{Co}{ii}  & 13.83$\pm$0.13 & --8.17$\pm$0.16 & --7.01 & --1.16 & 13.73$\pm$0.21 \\
\ion{Ni}{ii}  & 14.09$\pm$0.19 & --7.91$\pm$0.21 & --5.78 & --2.13 & 14.22$\pm$0.21\\
\ion{Zn}{ii}  & 13.75$\pm$0.15 & --8.25$\pm$0.18 & --7.44 & --0.81 & 13.64$\pm$0.04
\enddata
\tablenotetext{a}{Solar photosphere abundances of the neutral species
  of each element from \citet{asplund09}.}
\tablenotetext{b}{Blended with \ion{C}{ii}*~$\lambda$1335, assuming
  the two lines contribute equally to the measured $W_r$.}
\tablenotetext{c}{Derived from the sum of the ionization levels. Lines
are saturated, so a lower limit can be determined.}
\tablenotetext{d}{Determined from the single line \ion{Al}{ii} $\lambda$1670.}
\tablenotetext{e}{Contaminated by \ion{Si}{ii}*.}
\tablenotetext{f}{The dominant species is \ion{Ca}{iii}.}
\tablenotetext{g}{Wide wavelength range for \ion{Fe}{i} (1875--2984
  {\AA}). Hence the composite probe different regions, making the fit
  to the curve of growth bad, and values between 13.5 $<$ 
  log\,$N$(\ion{Fe}{i}) $<$ 14.5 are possible.}

\end{deluxetable}

\end{document}